\documentclass[review]{elsarticle}
\usepackage{balance}
\usepackage{algorithm}
\usepackage{algpseudocode}
\usepackage{caption}

\usepackage{hyperref}
\usepackage{float}
\usepackage{verbatim} 
\usepackage{apalike}
\restylefloat{figure}
\restylefloat{table}

\usepackage{xurl}

\journal{Expert Systems with Applications}
\newcommand{\review}[1]{\textcolor{black}{#1}}
\biboptions{authoryear}

\usepackage{amssymb}

\usepackage{xcolor}
\usepackage{todonotes}

\journal{Expert Systems with Application}

\begin{document}

\begin{frontmatter}

\title{Temporal Analysis of Drifting Hashtags in Textual Data Streams: A Graph-Based Application}

\author[inst1,inst2]{Cristiano Mesquita Garcia\corref{cor1}}
\ead{cristiano.garcia@ifsc.edu.br}

\author[inst1,inst3]{Alceu de Souza Britto Jr}
\ead{alceu@ppgia.pucpr.br}

\author[inst1]{Jean Paul Barddal}
\ead{jean.barddal@ppgia.pucpr.br}

\cortext[cor1]{Corresponding author.}
\address[inst1]{Programa de Pós-graduação em Informática (PPGIa), \\Pontifícia Universidade Católica do Paraná (PUCPR), Curitiba, Brazil}
\address[inst2]{Instituto Federal de Santa Catarina (IFSC), Caçador, Brazil}
\address[inst3]{Universidade Estadual de Ponta Grossa (UEPG), Ponta Grossa, Brazil}



            

\begin{abstract}

Initially supported by Twitter, hashtags are now \review{used} on several social media platforms.
Hashtags are helpful \review{for tagging, tracking, and grouping} posts on similar topics.
\review{In this paper, based on a hashtag stream regarding the hashtag \#mybodymychoice, we analyze hashtag drifts over time using concepts from graph analysis and textual data streams using the Girvan-Newman method to uncover hashtag communities in annual snapshots between 2018 and 2022.}
In addition, we offer insights about some correlated hashtags found in the study.
Our approach can be useful for monitoring changes over time in opinions and sentiment patterns about an entity on social media.
Even though the hashtag \#mybodymychoice was initially coupled with women's rights, abortion, and bodily autonomy, we observe that it suffered drifts during the studied period across topics such as drug legalization, vaccination, political protests, war, and civil rights.
The year 2021 was the most significant drifting year, in which the communities detected \review{and their respective sizes} suggest that \#mybodymychoice \review{had a significant} drift to vaccination and Covid-19-related topics.
\end{abstract}



\begin{keyword}
text drift \sep hashtag drift \sep temporal analysis \sep textual data streams
\end{keyword}

\end{frontmatter}


\section{Introduction}
\label{sec:intro}

The popularization of the Internet made it possible for people to express themselves in several ways via posts on social media using text, audio, videos, and pictures. 
People can also expose opinions about any topic, e.g., politics, products, and sports, and engage in collective actions and social movements.
Social media has also enabled social movements to take place worldwide. 
For instance, authors in \cite{eltantawy2011arab} discuss the role of social media in the Egyptian Revolution in 2011. 
The authors in \cite{harlow2012social} analyzed Guatemalan users' posts on Facebook groups to evaluate whether social media helped the movement in defense of Rodrigo Rosenberg to go offline in 2010. 
In addition, the authors in \cite{harlow2012social} named these digital movements `Activism 2.0', also due to Web 2.0, which includes social media networks. Another example of such activism in social media is the `Arab Spring,' also discussed in several papers \citep{smidi2017social,wolfsfeld2013social,comunello2012will}. 

More recently, during the Covid-19 pandemic, several movements in favor and against vaccination and other aspects of the pandemic, e.g., lockdown, have emerged \citep{sufi2022tracking}. 
In \cite{menghini2022drift}, the authors discussed the use of \textit{\#mybodymychoice}, a feminist slogan to defend women's rights, used for opposing Covid-19 vaccination campaigns. 

Twitter is one of the most popular social media platforms. 
Around 500 million tweets are posted per day in 2023\footnote{\url{https://www.omnicoreagency.com/twitter-statistics/}}. 
Each post is known as \textit{tweet}. 
Its ease of posting and sharing information made it popular also for famous people, such as artists, athletes, and politicians. 
As of 2017, Twitter limits the post to 280 characters in its free version. 
In addition, people can tag their posts by using \textit{hashtags}, leading to a post categorization. 
Researchers that need to evaluate information from \textit{social sensors}, e.g., public opinion, resort to data from social media \citep{suprem2019concept}.  


Inspired by the authors in \cite{menghini2022drift}, we also analyzed the Twitter stream regarding the hashtag \textit{\#mybodymychoice}, comprising the years between 2018 and 2022. 
The authors in \cite{menghini2022drift} demonstrated how the hashtag mentioned before, originally used to defend the right to abortion and other bodily autonomy aspects, drifted over time, being also used in different topics. 
Our work technically analyzes the context evolution of the hashtag, originating from a textual data stream corresponding to \textit{\#mybodymychoice} without judging the validity of the related social causes or the drifts of this hashtag. 
In addition, different from \cite{menghini2022drift}, which split the stream using two events known beforehand, we leveraged the Girvan-Newman community detection algorithm \citep{girvan2002community} between years to evaluate the communities found.

This paper aims to analyze the drifts in the hashtag stream of \textit{\#mybodymychoice} using time slices corresponding to years between 2018 and 2022, inclusive. 
We applied the Girvan-Newman algorithm for community detection after each year was finished. We analyzed the most relevant, i.e., the biggest, communities by evaluating the hashtags and their contexts. 
We are interested in analyzing the \textit{\#mybodymychoice}'s distancing from its original use or context. 
In addition, we listed the five most frequent hashtags per year. 
We consider drifts to have occurred whenever the hashtag is used in \review{contexts different from} its original. 
Therefore, our research questions are (RQ1) ``How \review{did} the hashtag \#mybodymychoice evolve from 2018 to 2022?'', and (RQ2) ``What were the most significant hashtag drifts over time, considering the graph communities detected?''. 
RQ1 aims to uncover drifts and their importance regarding a conceptual similarity to the original use of \textit{\#mybodymychoice}. 
RQ2 aims to find which topics other than women's rights, abortion, and bodily autonomy were frequent in the period analyzed. \review{Although detecting hashtag drifts has been addressed in the literature, most analyses indicate to be in batch, thus demanding all the data forehand. Therefore, our motivation is to provide an online method to analyze hashtag streams subject to hashtag drifts that could be directly coupled to a streaming source, such as a social media platform Application Programming Interface (API).}

The contribution of this work is \review{three}-fold: (i) the analysis of the hashtag stream respecting the textual data stream concepts and using graph-based approaches, including community detection between years, (ii) to the best of our knowledge, this is the first time the Girvan-Newman algorithm was applied to a hashtag graph generated in a streaming fashion\review{, and (c) an online graph management method to maintain the hashtag graph concise and with the most relevant hashtags}. A by-product of this work is the dataset containing hashtag stream regarding the \textit{\#mybodymychoice} hashtag between 2018 and 2022. 
This work may be relevant to applications that monitor sentiments or opinions regarding an entity on social media to recognize changes in patterns and behaviors. \review{The innovation of this work lies in the incremental graph generation method, which allows: (a) the management of the graph conciseness while preserving important hashtags, (b) the direct coupling to a text streaming API for online analysis, and (c) analyses over time, whenever it is needed.}


This work is organized as follows. Section \ref{sec:background} presents the context background, including hashtags, textual data stream mining, graphs, and community detection.
Section \ref{sec:methodology} describes the methodology used in the analysis, including the dataset collection and hashtags treatment. Section \ref{sec:exp-protocol} presents the graph generation and implementation. 
Section \ref{sec:results-analysis} analyzes the hashtags that co-occur with \textit{\#mybodymychoice} to infer topics using community detection in graphs, considering the years between 2018 and 2022. 
Finally, Section \ref{sec:conclusion} concludes this work and provides future directions.

\section{Background}
\label{sec:background}

This section introduces the concepts to understand the approach. We present hashtags and their uses, textual data stream mining concepts, graphs, and community detection in graphs.

\subsection{Hashtags}
\label{subsec:hashtags}

Hashtags are words starting with the hash character (\#) used for social media post categorization. Hashtags, under the name of \textit{channel tags}, were elaborated by Chris Messina in 2007 with the idea of creating groups on Twitter\footnote{Available at: \url{https://factoryjoe.com/2007/08/25/groups-for-twitter-or-a-proposal-for-twitter-tag-channels/}, accessed on June 20th, 2023}. 
Twitter adopted hashtags as a feature in 2009 \citep{scott2015pragmatics}, which proved helpful and, later, was also integrated into other social media platforms, such as Facebook\footnote{Available at: \url{https://about.fb.com/news/2013/06/public-conversations-on-facebook/}. Accessed on June 20th, 2023.} in 2013.

Twitter post categorization or grouping leads to virtual community creation. 
\review{The creation of virtual communities} enables discussions on several aspects, and sometimes, online social movements go offline \citep{smidi2017social}. 
An example of a hashtag used in social discussion is \textit{\#mybodymychoice}.
As pointed out in \cite{menghini2022drift}, the hashtag \textit{\#mybodymychoice} initially regarded women's right to abortion and bodily autonomy. 
However, the authors in \cite{menghini2022drift} found that between 2018 and 2021, this hashtag drifted to other topics, such as politics and civil rights.

This paper expands the analysis performed in \cite{menghini2022drift} to 2022, considering hashtags co-occurrence and community detection via temporal graphs and textual streaming analysis. In our analysis, we intend to trace back the events that used particular hashtags to understand the hashtags' occurrences.
Therefore, this section presents textual data stream mining and its characteristics, graphs and their notation, and techniques for community detection in graphs.


\subsection{Textual Data Stream Mining}
\label{subsec:tsdm}
Data streams are ``an algorithmic abstraction that allows for real-time analytics'' \citep{gama2014survey}. Data arrive sequentially (one by one or in small batches) and quickly in streaming settings. 
In addition, a data stream may be infinite and ephemeral, meaning that the data distribution may change over time, thus giving rise to a phenomenon named concept drift \citep{MOA-Book-2018,gama2014survey}. 

Therefore, to learn from data streams, i.e., data stream mining, some capabilities desired for machine learning methods include \citep{MOA-Book-2018,gama2014survey}: (a) learning quickly from the data as it arrives, (b) discarding data within a short time after learning from it, (c) performing single-pass processes, and (d) detecting and adapting to concept drifts.

Textual data streams are specializations of traditional data streams \citep{thuma2023benchmarking}, i.e., \review{they comprise} a sequence of texts arriving continuously at a fast pace. 
However, processing texts in textual data stream scenarios is more challenging given the natural language processing (NLP)-related activities regarding, for example, vocabulary maintenance, token standardization, and processing texts only once.

Our proposed approach focuses on processing textual data streams while accounting for the constraints \review{mentioned above}. 
The exception regards graph community detection, detailed in Section \ref{subsec:cd-g}, which is performed on a year-basis and, thus, should be considered an \emph{offline step}.

\subsection{Graph}
\label{subsec:gt}

\review{As stated in \cite{bondy1976graph}, ``many real-world situations can conveniently be described by means of a diagram consisting of a set of points together with lines joining certain pairs of these points''. As examples, these points and lines could represent cities and the roads between them, people and their relationships, or companies and trades between them. The representation based on points and lines is called \textit{graph}.}

\begin{figure}[!htp]
    \centering
    \includegraphics[width=.55\columnwidth]{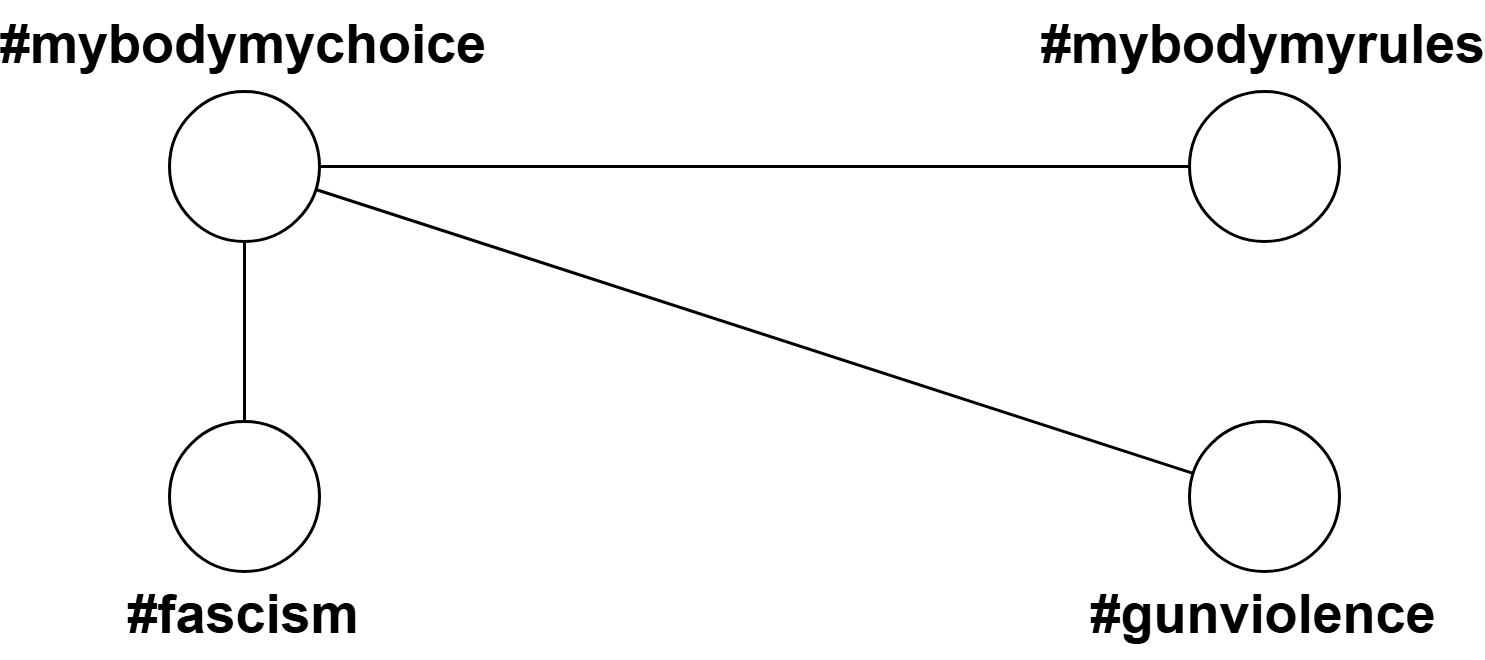}
    \caption{An example graph with four vertices (\#hashtags) and three undirected edges (lines).}
    \label{fig:graph}
\end{figure}

A graph $G$ is a collection of \review{finite} edges $E$ and \review{finite} vertices $V$\review{, generally being represented as $G(V,E)$}. Each vertex can be linked to another vertex by an edge. Generally, vertices are visually represented as circles, and the edges are lines between vertices, as depicted in the example graph in Fig. \ref{fig:graph}. 
Edges are directed if the direction between vertices matters and are undirected otherwise. 
Edges may also be coupled with weights to denote additional information about the relationship between connected vertices. 

Graphs are relevant to denote entities and \review{their relationships, which are} very useful in text analysis. 
For instance, to find relevant words in sentences, the vertices are the words, and the edges are their relationships \citep{castillo2017text}. 
In our proposal, vertices denote hashtags, and edges \review{indicate} the co-occurrence between the linked hashtags. 
The edges are not weighted.

\subsection{Community Detection in Graphs}
\label{subsec:cd-g}

\review{Community detection methods aim at identifying groups with frequent links, indicating those groups as communities \citep{daud2020applications}.}
Criminal community detection \citep{sarvari2014constructing}, cattle trade movement \citep{cardosostructure}, and disease spread dynamics \citep{salathe2010dynamics} are applications that leverage community detection. 
Several community detection methods are available in the literature, such as the Louvain method \citep{blondel2008fast}, the Leiden algorithm \citep{traag2019louvain}, and the Girvan-Newman algorithm \citep{girvan2002community}.

Similar to clustering, community detection demands the graph partition into densely connected nodes and nodes from a \review{particular} community to be far from those of other communities \citep{blondel2008fast}. 
According to \cite{traag2019louvain}, the Louvain method obtains weakly connected communities. 
To handle this characteristic, the authors in \cite{traag2019louvain} added one more phase to the Louvain algorithm to obtain better-connected communities.

The Girvan-Newman algorithm \citep{girvan2002community} for community detection on graphs performs an iterative edge elimination. 
To consider an edge for elimination, it must be part of the highest number of shortest paths between nodes, i.e., betweenness centrality. 
After removing these edges, the communities are obtained. 
More specifically, the algorithm works in four steps \citep{girvan2002community}: (a) for all edges in the graph, compute the betweenness; (b) remove the edge with the highest betweenness; (c) for all edges affected by the edge removal in step (b), re-compute the betweenness; (d) return to step (b) until there is no edge remaining.
We resorted to the Girvan-Newman algorithm because it generates consistent communities \citep{sheng2020research}. 
However, its trade-off lies in its complexity, which makes it not scalable for large networks \citep{sheng2020research}. 

Considering that we are using a textual data stream limited by a window and that the communities are detected in an offline step, community quality is the most important. 
Therefore, the \review{complexity of the Girvan-Newman algorithm} is not a concern in this particular scenario. 

\subsection{\review{Related Works}}

\review{As mentioned in Section \ref{subsec:hashtags}, hashtags have been used in social media platforms as a form of post categorization. However, \textit{hashtag drift} is a recently studied phenomenon. Sometimes, this phenomenon is referred to as \textit{hashtag hijacking} \citep{vandam2016detecting} or \textit{hashjacking} \citep{darius2019twitter}. Below, we present some papers that consider hashtag drift and applications.}

\review{The authors in \cite{booten2016hashtag} described the hashtag drift based on evolving uses of political hashtags. The authors divided the hashtags into \textit{focused} and \textit{individualistic}. Focused hashtags are those which have their use still in the original context, while individualistic hashtags are the hashtags that evolved according to individual's particular contexts. In this paper, the authors used Tumblr API as the data source. The authors filtered through the API to obtain posts using the most important hashtag in the past years according to a set of web pages. In their experiments, the authors kept nine hashtags, since more than 5,000 posts included at least one of these hashtags. The authors obtained the earliest and latest deciles of posts for each hashtag. The authors also proposed a score to evaluate drifts, considering their observation that states \textit{focused hashtags} tend to have a small set of frequent co-occurring hashtags, while \textit{individualistic hashtags} have a ``great diversity of infrequent co-occurring hashtags''. The authors conclude by confirming a finding of \cite{bennett2011digital}, which claimed that ``activist movements can survive and thrive by allowing individuals to flexibly re-define the movement's goals on social media'' \citep{bennett2011digital,booten2016hashtag}.}

\review{Concerned about the challenges of antibiotic resistance, the authors in \cite{mackenzie2020world} evaluated the impact of particular hashtags on disseminating information regarding the overlapping campaigns European Antibiotic Awareness Day and World Antibiotic Awareness Week in 2018. The authors considered a set of 14,400 tweets with 60,222 retweets. In this paper, the authors identified hashtag drift. However, they use a slightly different definition, i.e., ``the use and spread of hashtags other than the official hashtags''. One of the findings of the authors stated that an unofficial hashtag \#WAAW2018 was a strong predictor of retweet activity for users with less than 1,000 followers.}

\review{The authors in \cite{arif2018acting} analyzed discourses regarding the \#BlackLivesMatter between January and October 2016. This hashtag had its origin in response to the acquittal of George Zimmerman after the shooting death of Trayvon Martin, which happened in 2012. The authors' filtering resulted in a dataset with 248,719 tweets. When visualizing their dataset, the authors could label them according to the political spectrum. Among their findings, the authors realized a phenomenon close to hashtag drift. For example, some tweets tried to impose new meanings to \#BlackLivesMatter, such as ``adversarial stance towards law enforcement'' \citep{arif2018acting}.}

\review{In \cite{menghini2022drift}, the authors analyzed the drift of \#MyBodyMyChoice on Twitter, using tweets from January 2018 to December 2021. The authors split the period into three moments, having the Covid-19 pandemic as the reference, i.e., before the pandemic (BP), initial pandemic (IC), and Covid-19 coexistence (CC). The authors noticed a drift/hashjacking related to Covid-19 and vaccination from when the first official Covid-19 case outside China was confirmed. Compared to this case specifically, we evaluated between 2018 and 2022 (inclusive) using the stream paradigm. Herefore, the approach could be directly coupled to a streaming source, such as a social media platform API. In addition, we evaluated the co-occurrences of hashtags to verify drift other than those related to Covid-19.}

\review{More recently, the authors in \cite{fitzgerald2024savethechildren} evaluated the use of \#SaveTheChildren between January 2022 and March 2023 on Twitter (X). The authors assessed that the aforementioned hashtag was used in contexts different from the original. According to the authors, \#SaveTheChildren is often associated with charity to aid children and posts related to Covid-19 vaccines. However, the authors also verified that this hashtag is used to spread conspiracy theories related to anti-vaccination and anti-LGBTQIA+ rhetoric. In addition, the authors checked that conspiratorial tweets are more retweeted than non-conspiratorial tweets. In addition, the authors highlight that drifted hashtags can impact legitimate movements, and these drifted/hijacked hashtags can expose more users to conspiracy theories.}

\review{As seen, hashtag drift/hijacking, or hashjacking, was identified and has been studied under several aspects, such as political and technical. However, the main difference in our approach is the online construction and maintenance of a hashtag co-occurrence graph with yearly analysis. This characteristic allows the online analysis using a streaming data source. Therefore, with the huge data production and big data scenarios, our approach can be relevant.}

\section{Methodology}
\label{sec:methodology}

In this section, the data collection is detailed \review{and} discussed, and the obtained dataset is described. 
In addition, we present the treatment \review{of hashtags}.


\subsection{Dataset description}

\begin{figure}[!htp]
    \centering
    \includegraphics[width=\columnwidth]{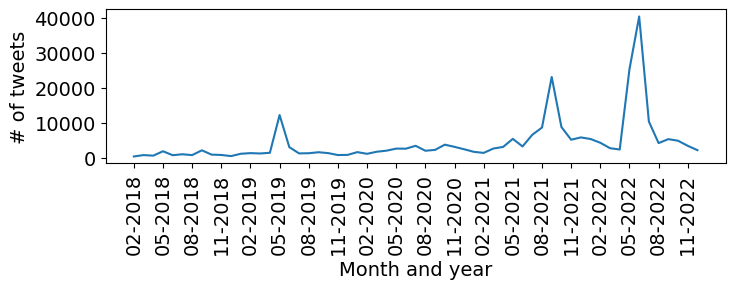}
    \caption{Number of tweets distributed across the period evaluated.}
    \label{fig:tweets-distribution}
\end{figure}

The dataset regarding the \textit{\#mybodymychoice} hashtag comprises tweets from February 11th, 2018, to December 31st, 2022. 
Data was acquired using the Twitter API\footnote{\url{https://developer.twitter.com/en/docs/twitter-api}} with the ``$(\#mybodymychoice)\ -is:retweet$'' query, which ignores retweets (re-posts). 
In addition, we performed the collection respecting Twitter's policies for that. 
We ignored retweets since they could bias our analysis by inflating the use of particular hashtags. 
The dataset totaled 255,131 valid tweets. 
Figure \ref{fig:tweets-distribution} depicts the tweet distribution over the analyzed period.


The peaks in Figure \ref{fig:tweets-distribution} provide interesting insights.
To provide information to investigate the peaks, we collected the five most frequent hashtags in the months of interest, i.e., May 2019, August and September 2021, and May 2022.
In the first peak, i.e., May 2019, the five most frequent hashtags, together with their respective frequencies, were \#prochoice (1090), \#womensrightarehumansrights (863), \#womensrights (728), \#abortionisawomansright (713), and \#stopthebans (552). 
Investigating this peak, we retrieved that in May 2019, several protests against abortion-related restrictions happened in the United States\footnote{Available at: \url{https://www.npr.org/2019/05/21/725410050/across-the-country-protesters-rally-to-stop-states-abortion-bans}. Accessed on May 30th, 2023.}.

Repeating the analysis for the second peak in August 2021, the five most frequent hashtags and respective frequencies were: \#novaccinepassports (519), \#freedom (443), \#novaccinepassportsanywhere (397), \#novaccinepassport (382), and \#covid19 (330). We can see that this peak starts earlier than August. It matches the release of the European Union Digital Covid Certificate, released on July 1st, 2021\footnote{Available at: \url{https://commission.europa.eu/strategy-and-policy/coronavirus-response/safe-covid-19-vaccines-europeans/eu-digital-covid-certificate\_en}. Accessed on May 30th, 2023.}. 
In September, the most frequent hashtags were \#womenrights (1321), \#texaswaronwomen (1282), \#abortionishealthcare (1268), \#prochoice (1231), and \#novaccinepassports (1187).
The second hashtag provides interesting information about the ``Texas war on women''. 
In Texas, protests against the state abortion-related laws emerged in this period\footnote{Available at: \url{https://www.theguardian.com/global-development/2021/sep/18/texas-anti-abortion-law-shows-terrifying-fragility-of-womens-rights-say-activists}. Accessed on May 30th, 2023.}.

The third peak, between May and August 2022, reflects a significant event regarding the \textit{\#mybodymychoice}. 
In June 2022, the US Supreme Court invalidated the Roe vs. Wade decision (which favored abortion in 1973)\footnote{Available at: \url{https://www.supremecourt.gov/opinions/21pdf/19-1392\_6j37.pdf}. Accessed on May 30th, 2023.}. 

In summary, the data collected provides several insights and reflects the past regarding the hashtag \textit{\#mybodymychoice}.
\review{Under Twitter's policies, the data and files with the generated graphs are publicly available at \url{https://github.com/cristianomg10/temporal-analysis-of-drifting-hashtags-in-textual-data-streams-a-graph-based-application}}. 

\subsection{\review{Treatment of} Hashtags}

This work verifies the relationships of the hashtag \textit{\#mybodymychoice} over time. 
Therefore, the texts were discarded, links removed, and we leveraged only the hashtags in each tweet.
Hashtags were converted into lowercase, and those with less than three characters were discarded.
In addition, we removed special characters such as `;', `!', and `$\backslash$n'.

\subsection{Evaluation}

\review{The evaluation of the approach is performed qualitatively. This means that the analysis is performed after each year finishes (according to the tweets' timestamps). The most extended communities have their nodes, i.e., hashtags, analyzed in order to find patterns and infer the context.}

\review{Since we analyze the hashtag \#mybodymychoice in this paper, we consider drifts to have occurred whenever the hashtag is used in contexts different from its original. That is, we consider drift if a community's context is different from feminism and bodily autonomy. In addition, in case of a drift occurrence, we consider the size of the community as the drift importance.}

\section{Experimental Protocol}
\label{sec:exp-protocol}
This section presents the graph generation process and its implementation, including the pseudocodes.

\subsection{Graph generation}
We generated graphs using their intrinsic characteristics: each node corresponded to a hashtag, and each edge between two nodes corresponded to their co-occurrence. 
We considered a hashtag valid once it appears in the dataset at least five times to ease the analyses and avoid noise. 
\review{We perform the computation on the fly considering the textual data stream setting. }
\review{This} means that the graph generation process respects the \review{constraints of} the textual data stream. 
The only exception is the community detection performed at the end of each year (regarding the timestamp of the textual data stream).

In addition, to limit our analysis to the most recent and important hashtags, we used a 200-size window and a hashtag aging mechanism. 
In our approach, each hashtag has an age attribute. 
This attribute controls the hashtags to be considered in the analysis. 
Whenever a new hashtag is to be inserted, if the window is full, the oldest hashtag in the window is removed to accommodate the \review{latest} hashtag. 
If the window is not full, the hashtag is added. 
In each iteration, the hashtags' age \review{increases}, except for the recently received hashtag.

\subsection{Implementation}

We implemented the code in Python 3.9. 
We plotted the graphs with the help of the Pyvis\footnote{\url{https://pyvis.readthedocs.io/en/latest/}} and NetworkX libraries\footnote{\url{https://networkx.org/}}. 
Algorithm \ref{alg:tdsp} depicts the pseudocode. 
The algorithm receives a textual data stream $TS$ and creates a new Graph with a window size of 200 and a minimum frequency of five. 
Line 2 starts the stream iteration; line 3 obtains the following text (tweet); line 4 extracts a list of hashtags, which is sent to the \texttt{add\_node} graph's method, detailed in Algorithm \ref{alg:add-node}. 

\begin{algorithm}
\caption{Textual Data Stream Processing}\label{alg:tdsp}
\begin{algorithmic}[1]
\Require Textual data stream $TS$
\State $g \gets new\ Graph(200, 5)$
\While{$TS \neq \O $}
    \State $text \gets TS.get\_next\_text()$
    \State $hashtags \gets text.get\_hashtags()$
    \State $g.add\_node(hashtags)$    \Comment{Algorithm \ref{alg:add-node}}
\EndWhile
\end{algorithmic}
\end{algorithm}

The Algorithm \ref{alg:add-node} details the \texttt{add\_node} method. 
In line 1, all the hashtags' age is incremented. 
In line 3, for each hashtag in the received list, we check if the hashtag is not in the vocabulary (\texttt{vocab}), i.e., it is not in the graph, if it is in the \texttt{pregraph}, and if its frequency reached five simultaneously. 
\texttt{Pregraph} is a graph's internal structure that stores the hashtags until they reach the minimum defined frequency. 
Also, \texttt{g.vocab} is the structure in which we control the existence of nodes, i.e., \texttt{hashtags}. 
Once a given hashtag reaches the minimum frequency, the graph accommodates it in line 4. 
In line 5, the hashtag is removed from the \texttt{pregraph}. If the condition in line 3 is not met, it is checked if the hashtag is in the \texttt{pregraph}. 
If the condition is met, its frequency is incremented. Next, we evaluate the remaining hashtags as connections. 
In line 10, we call the variable \texttt{other\_hashtags}. 
The same process of checking existence in \texttt{g.vocab} and \texttt{pregraph} is performed between lines 12 and 18. 
In line 19, we connect the \texttt{hashtag} and the \texttt{other\_hashtag}. 
In lines 20 and 21, we reset the age of \texttt{hashtag} and \texttt{other\_hashtag}. 
The age reset is crucial so the window can identify and remove stale hashtags when opportune.

\begin{algorithm}
\caption{Add node mechanism}\label{alg:add-node}
\begin{algorithmic}[1]
\Require Hashtags list $hashtags$
\Require Graph $g$
\State $g.grow\_old()$
\For{each $hashtag$ \textbf{in} $hashtags$}
    \If{$hashtag \notin g.vocab$ \textbf{and} $hashtag \in g.pregraph$ \textbf{and} $g.pregraph(hashtag) \ge 5$}
        \State $g.create\_node(hashtag)$
        \State $g.pregraph.remove(hashtag)$
    \ElsIf{$hashtag \in g.pregraph$}
        \State $g.pregraph(hashtag).increment()$
        \State \textbf{continue}
    \EndIf
    \State $other\_hashtags \gets hashtags - hashtag$
    \For{each $co\_hashtag$ \textbf{in} $other\_hashtags$}
        \If{$co\_hashtag \notin g.vocab$ \textbf{and} $co\_hashtag \in g.pregraph$ \textbf{and} $g.pregraph(co\_hashtag) \ge 5$}
            \State $g.create\_node(co\_hashtag)$
            \State $g.pregraph.remove(co\_hashtag)$
        \ElsIf{$co\_hashtag \in g.pregraph$}
            \State $g.pregraph(co\_hashtag).increment()$
            \State \textbf{continue}
        \EndIf

        \State $hashtag.add\_connection(other\_hashtag)$
        \State $hashtag.age \gets 0$
        \State $other\_hashtag.age \gets 0$
    \EndFor
\EndFor
\end{algorithmic}
\end{algorithm}

We do not include the hashtag \textit{\#mybodymychoice} because all the nodes would \review{connect} to the \textit{\#mybodymychoice}'s node\review{, hampering the visualization and including an unnecessary overhead to the algorithm}.

\review{The computational complexity of our approach is $O(SH^2)$, where $H$ is the hashtags of a given tweet (post), and $S$ represents the stream. We consider the $H^2$ component the complexity of the relationship mapping between the hashtags of a tweet. Please notice that we use the text stream paradigm. Therefore, we process tweet by tweet. In this point of view, the complexity is $O(H^2)$.}

\subsection{\review{Software and Hardware specification}}
\review{The experiment and analysis described in this paper were performed using Python\footnote{\url{https://www.python.org/}}, version 3.9.12. In addition, we used Pyvis\footnote{\url{https://pyvis.readthedocs.io/en/latest/}} for graph visualization and NetworkX\footnote{\url{https://networkx.org/}} for the implementation of the Girvan-Newman method. We also used Matplotlib\footnote{\url{https://matplotlib.org/}} and Seaborn\footnote{\url{https://seaborn.pydata.org/}} to generate some of the figures for this paper. The experiment and analysis were run on a laptop with a 12th-generation Intel 2.3GHz and 64 GB of RAM on Windows 11.}





\section{Results}
\label{sec:results-analysis}

This section shows the results obtained sliced by year. 
As mentioned, we considered the most relevant communities for analysis.
It is crucial to highlight that the graphs are not reset between years: they are constructed and updated incrementally. 
However, the colors have no correspondence across the years. 
For example, the cluster in green in 2018 may not be the same in green in 2019.

\subsection{Analysis of 2018}

\begin{figure}[!h]
    \centering
    \includegraphics[width=\columnwidth]{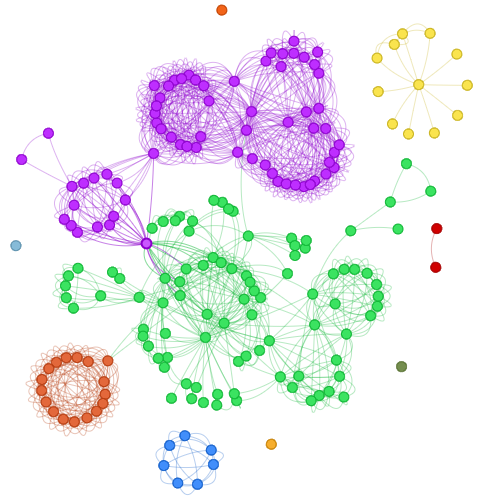}
    \caption{Relationship between hashtags, considering their co-occurrences. The colors represent the communities generated by the Girvan-Newman method. Best viewed in color.}
    \label{fig:2018}
\end{figure}

Figure \ref{fig:2018} depicts the remaining hashtags after the process described in Section \ref{sec:exp-protocol}. 
Each color represents a community detected by the Girvan-Newman method.

\begin{figure}[!h]
    \centering
    \includegraphics[width=\columnwidth]{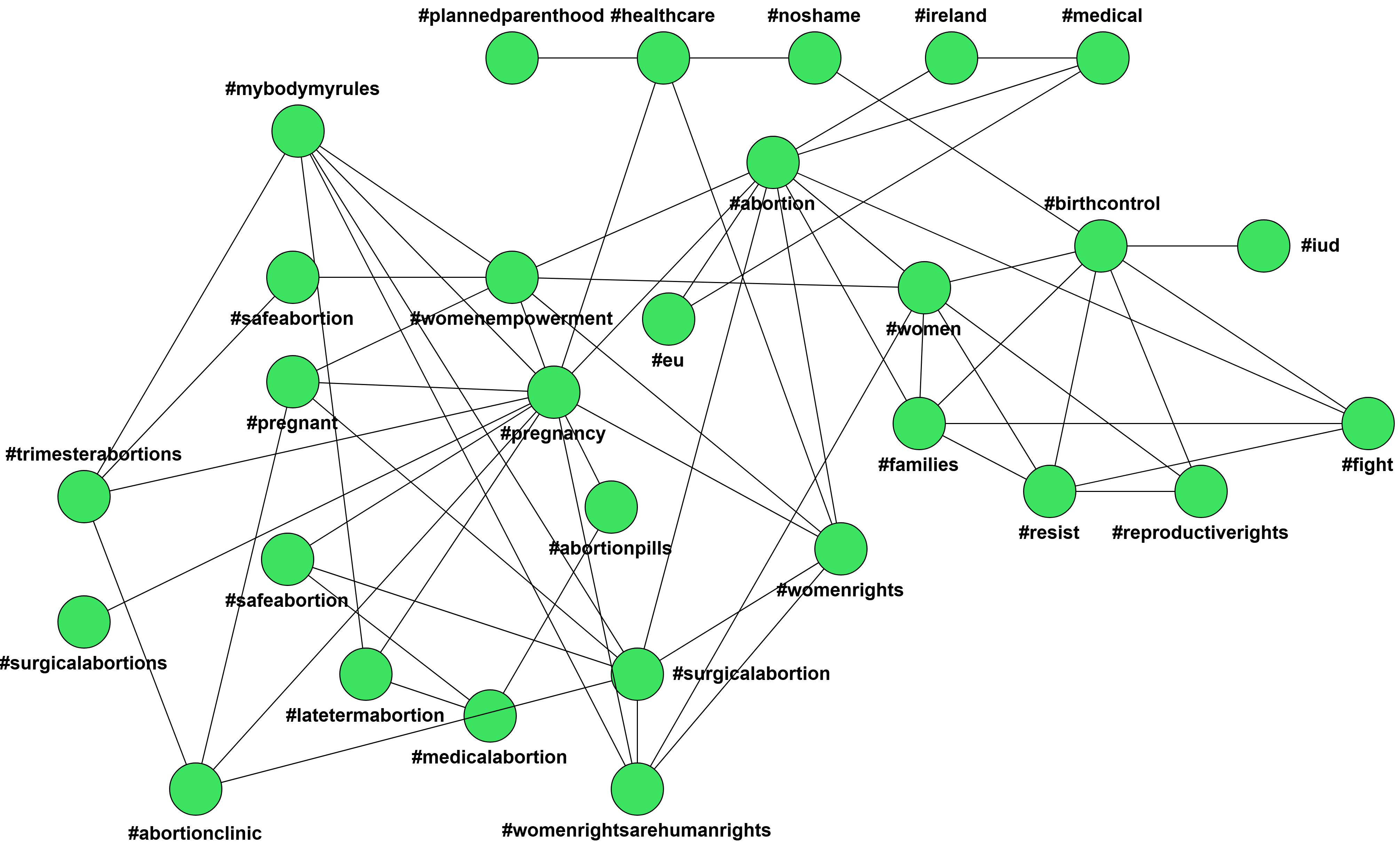}
    \caption{Part\review{ial view} of the green community from 2018.}
    \label{fig:zoom1-2018}
\end{figure}

\begin{figure}[!h]
    \centering
    \includegraphics[width=.7\columnwidth]{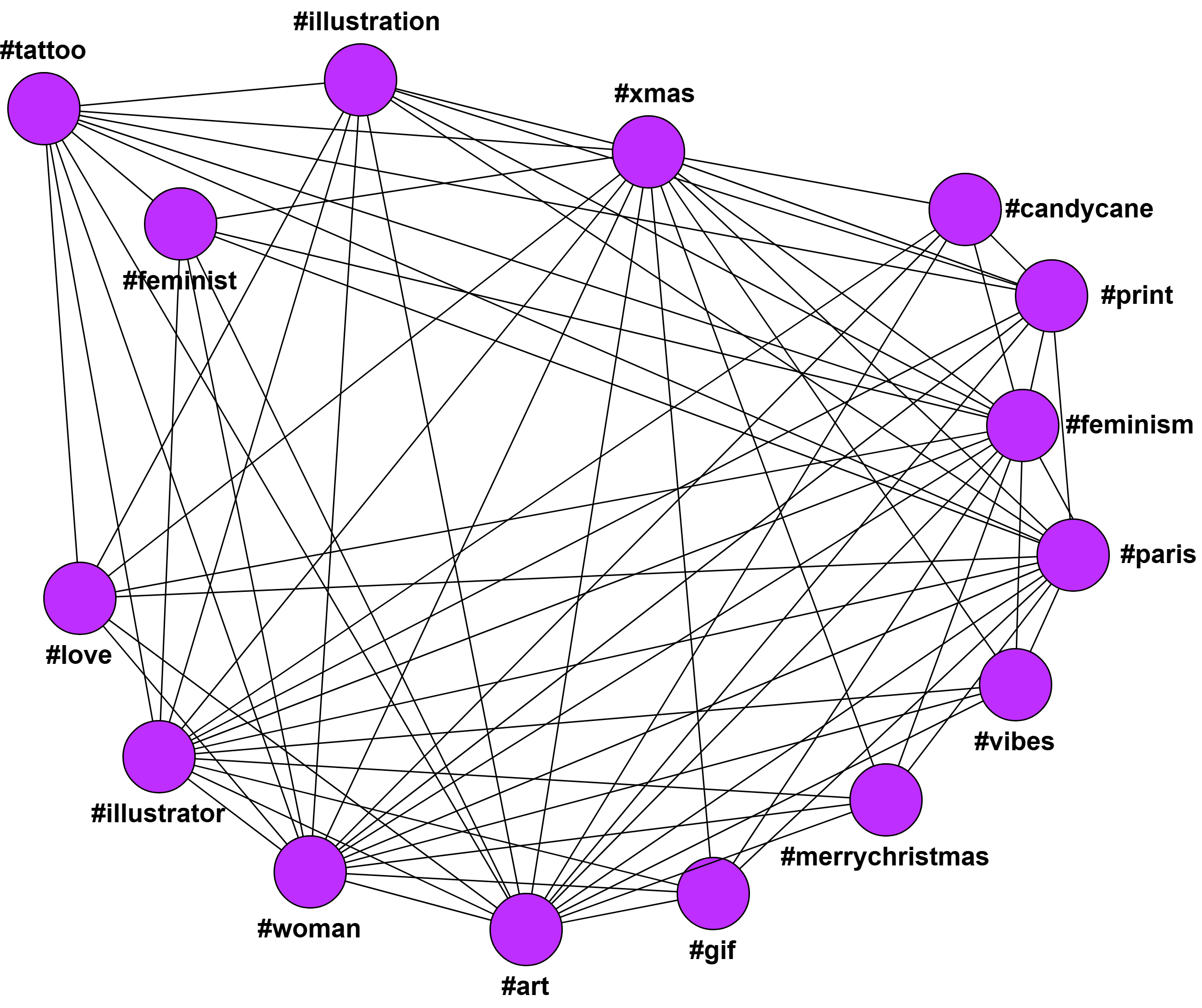}
    \caption{Part\review{ial view} of the purple community from 2018.}
    \label{fig:zoom2-2018}
\end{figure}

The largest community, i.e., in green, regards hashtags such as \#safeabortion, \#healthcare, \#reprorights, \#birthcontrol, and \#plannedparenthood. Therefore, this community is closely related to its original context: feminism and bodily autonomy, also considering abortion aspects. Fig. \ref{fig:zoom1-2018} partially shows the green community zoomed in. The community in purple regards different aspects, such as \#feminism, \#vibes, \#tattoo, and \#gaypower. The terms are mixed, and it is impossible to infer an exact context. Fig. \ref{fig:zoom2-2018} partially depicts this community.

\begin{figure}[!h]
    \centering
    \includegraphics[width=.7\columnwidth]{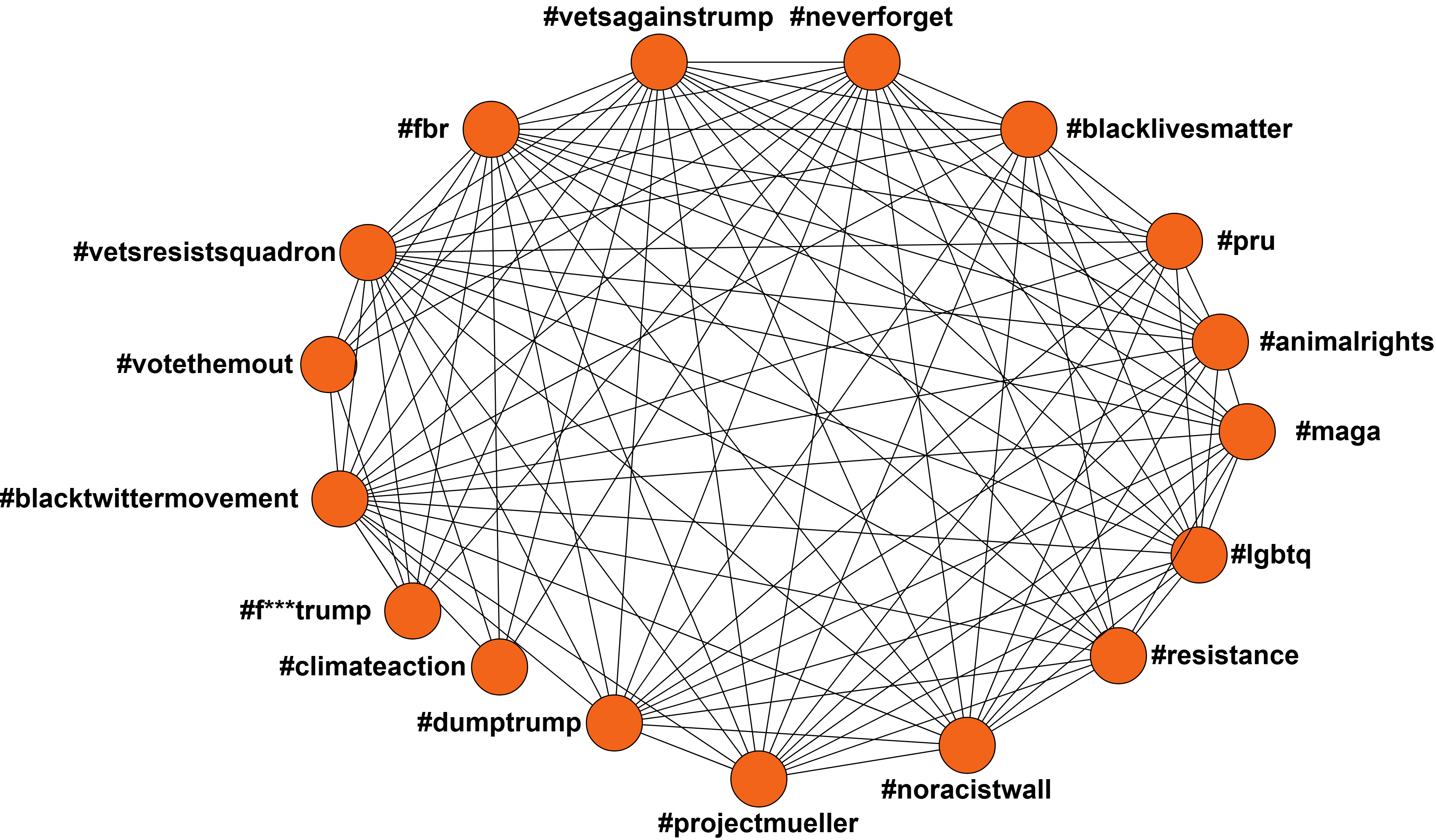}
    \caption{The orange community from 2018.}
    \label{fig:zoom4-2018}
\end{figure}

\begin{figure}[!h]
    \centering
    \includegraphics[width=.75\columnwidth]{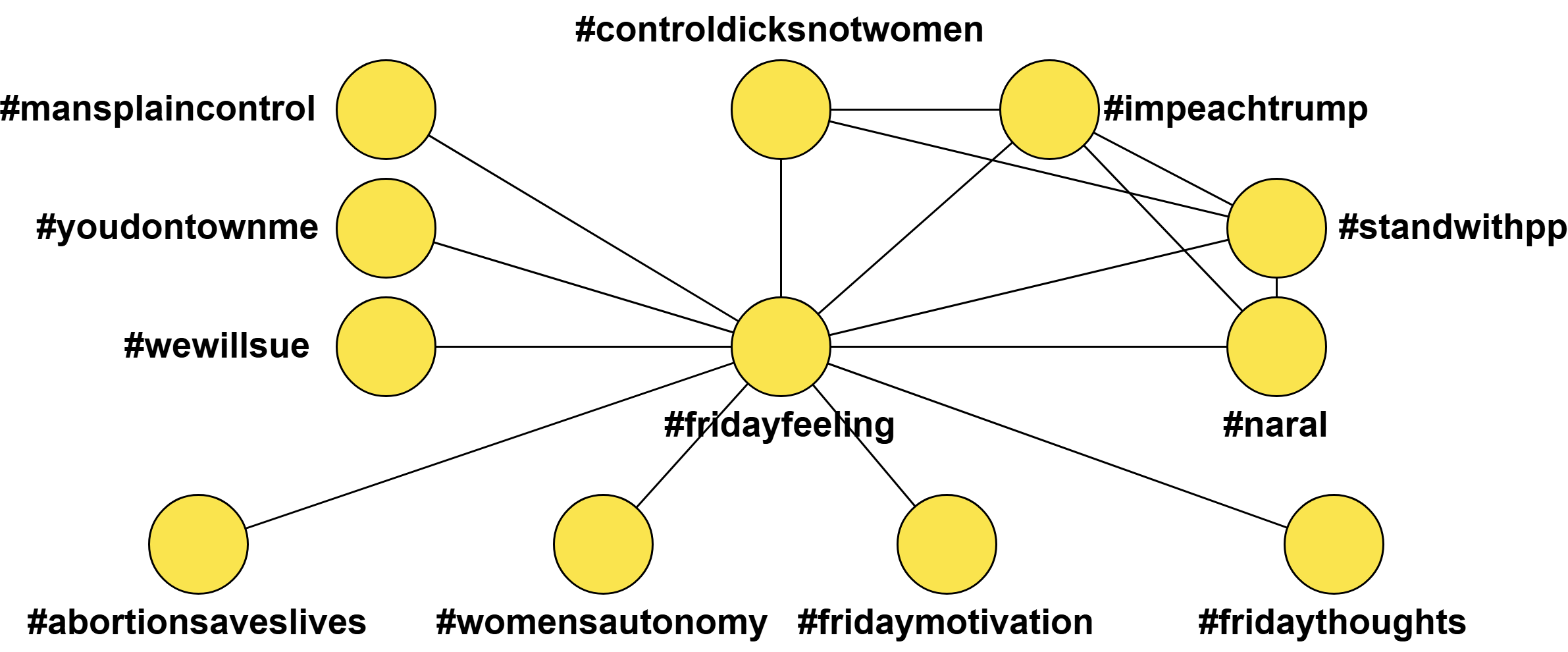}
    \caption{Yellow community from 2018.}
    \label{fig:zoom3-2018}
\end{figure}

Conversely, the community in orange, i.e., the community at the bottom left, considers hashtags that refer to civil rights, for example, \#blacklivesmatter, \#noracistwall, \#resistance, and \#votethemout. The community in yellow regards hashtags such as \#youdontownme, \#fridaythoughts, and \#impeachtrump. Although they are closely connected, inferring a context is impossible. Figs. \ref{fig:zoom4-2018} and \ref{fig:zoom3-2018} show the communities in orange and yellow, respectively. Some labels are not shown due to the existence of swear words.

\begin{figure}[!h]
    \centering
    \includegraphics[width=.55\columnwidth]{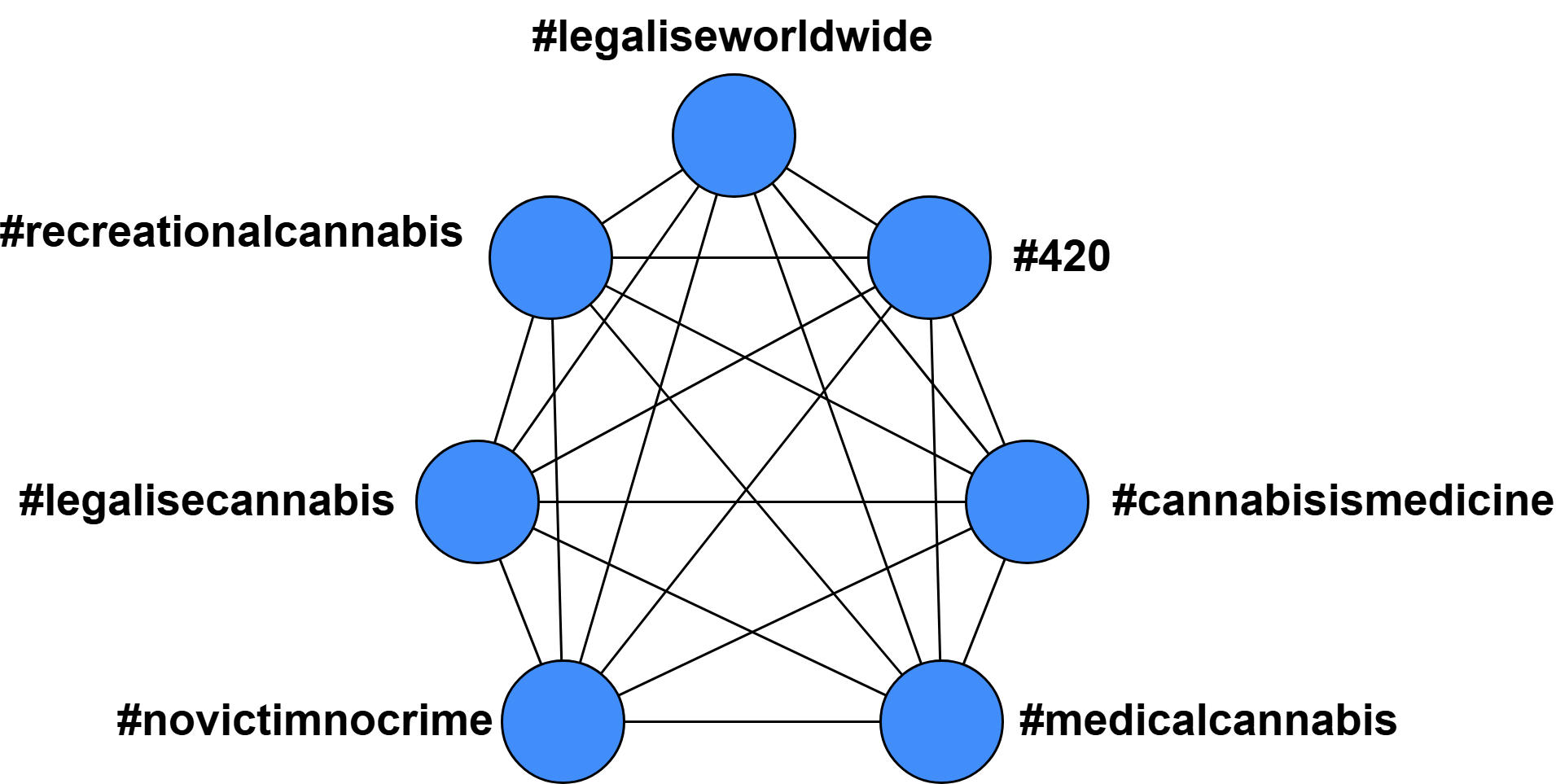}
    \caption{Blue community from 2018.}
    \label{fig:zoom5-2018}
\end{figure}

The community in blue considers the use of drugs. The hashtags that prevail are: \#novictimnocrime, \#recreationalcannabis, \#legalisecannabis, and \#cannabisismedicine. Fig. \ref{fig:zoom5-2018} depicts the blue community.

In summary, regarding 2018, the hashtag \textit{\#mybodymychoice} was used in different contexts than its original one, drifting to civil rights, the use of drugs, and politics. However, the largest community regards the original context: feminism, and women's bodily autonomy. The five most frequent hashtags in 2018 and respective frequencies were: \#prochoice (987), \#repealthe8th (532), \#womensrights (482), \#metoo (379), and \#fbr (339).
\review{Table \ref{tab:node_colors2018} summarizes the top 5 communities detected at the end of 2018 sorted by size, with their respective hashtag counts.}

\begin{table}[h!]
\centering
\begin{tabular}{cc}
\hline
\textbf{Context} & \textbf{\# of Hashtags}\\ \hline
{Feminism and bodily autonomy} & 90\\
{Mixed} & 68  \\ 
{Civil rights} & 17  \\
{Mixed} & 12  \\ 
{Drug liberation} & 7  \\ 
\hline 
\end{tabular}
\caption{Top 5 communities and their respective contexts and number of hashtags, for 2018.}
\label{tab:node_colors2018}
\end{table}

\subsection{Analysis of 2019}

\begin{figure}[!h]
    \centering
    \includegraphics[width=\columnwidth]{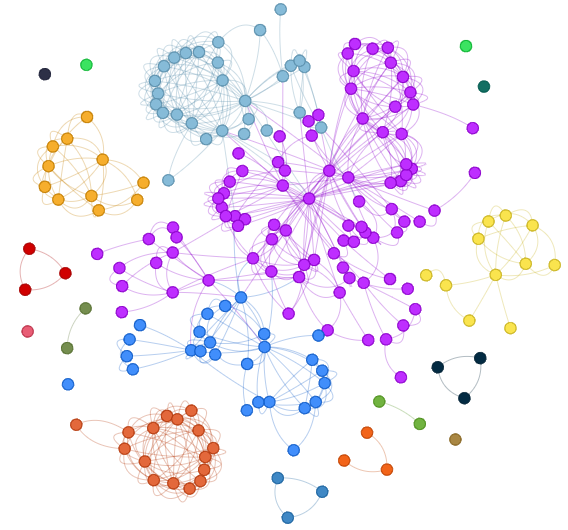}
    \caption{Relationship between hashtags, considering their co-occurrences regarding 2019. The colors represent the communities generated by the Girvan-Newman method. Best viewed in color.}
    \label{fig:2019}
\end{figure}

\begin{figure}[!h]
    \centering
    \includegraphics[width=.95\columnwidth]{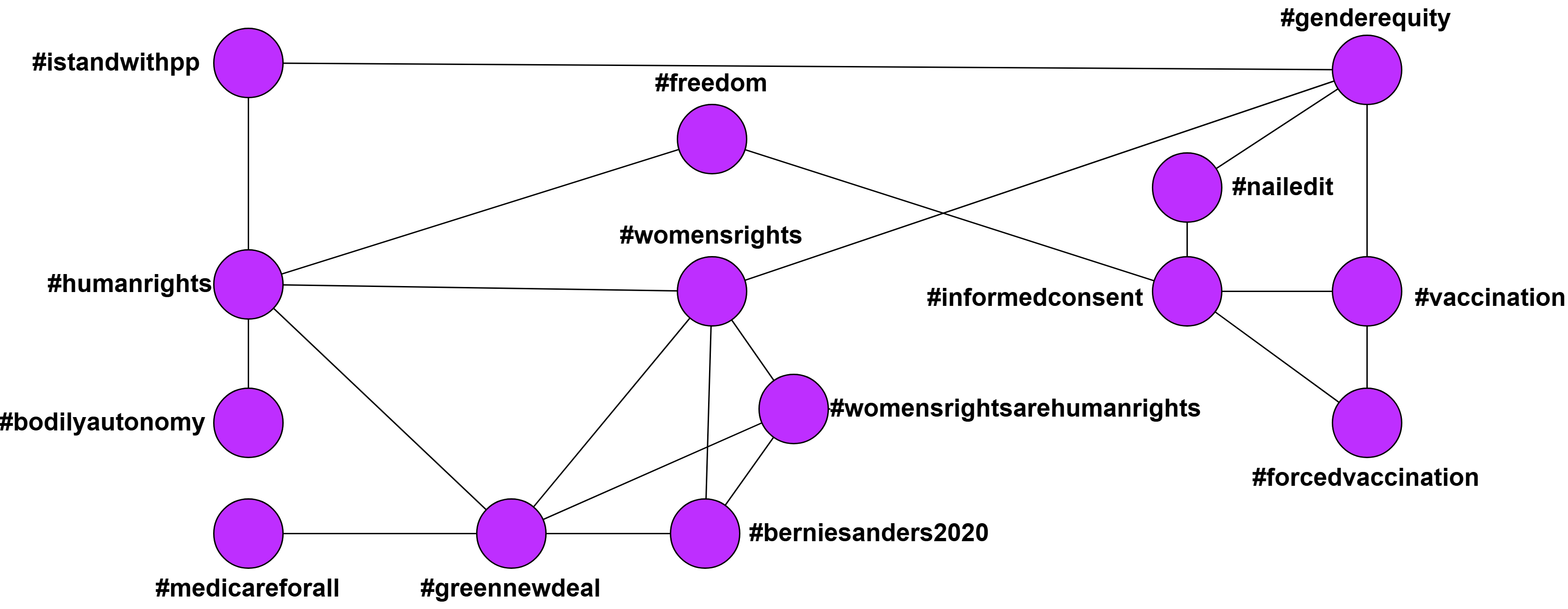}
    \caption{Part\review{ial view} of the purple community from 2019.}
    \label{fig:zoom1-2019}
\end{figure}

Fig. \ref{fig:2019} shows the communities detected by the Girvan-Newman method from the graph updated until the end of 2019. 
The biggest community is the purple one, partially depicted in Fig. \ref{fig:zoom1-2019}. This community includes hashtags across different topics such as \#humanrights, \#vaccination, \#forcedvaccination, \#drugs, \#prochoice, \#climateemergency, \#greennewdeal, and \#presidentberniesanders. Therefore, at first sight, it ranges from human rights to politics. According to his official website\footnote{Available at: \url{https://berniesanders.com/issues/reproductive-justice-all/}. Accessed on May 27th, 2023}, Bernie Sanders considers abortion a constitutional right. Therefore, it makes sense that the hashtag \#presidentberniesanders and other hashtags which seem to be related to Bernie Sanders' campaign for president in the United States appear together with \#prochoice and \#humanrights. In addition, \#greennewdeal is a topic of his campaign related to environmental issues. In essence, this community regards the civil rights and campaign topics of Bernie Sanders.

\begin{figure}[!h]
    \centering
    \includegraphics[width=.95\columnwidth]{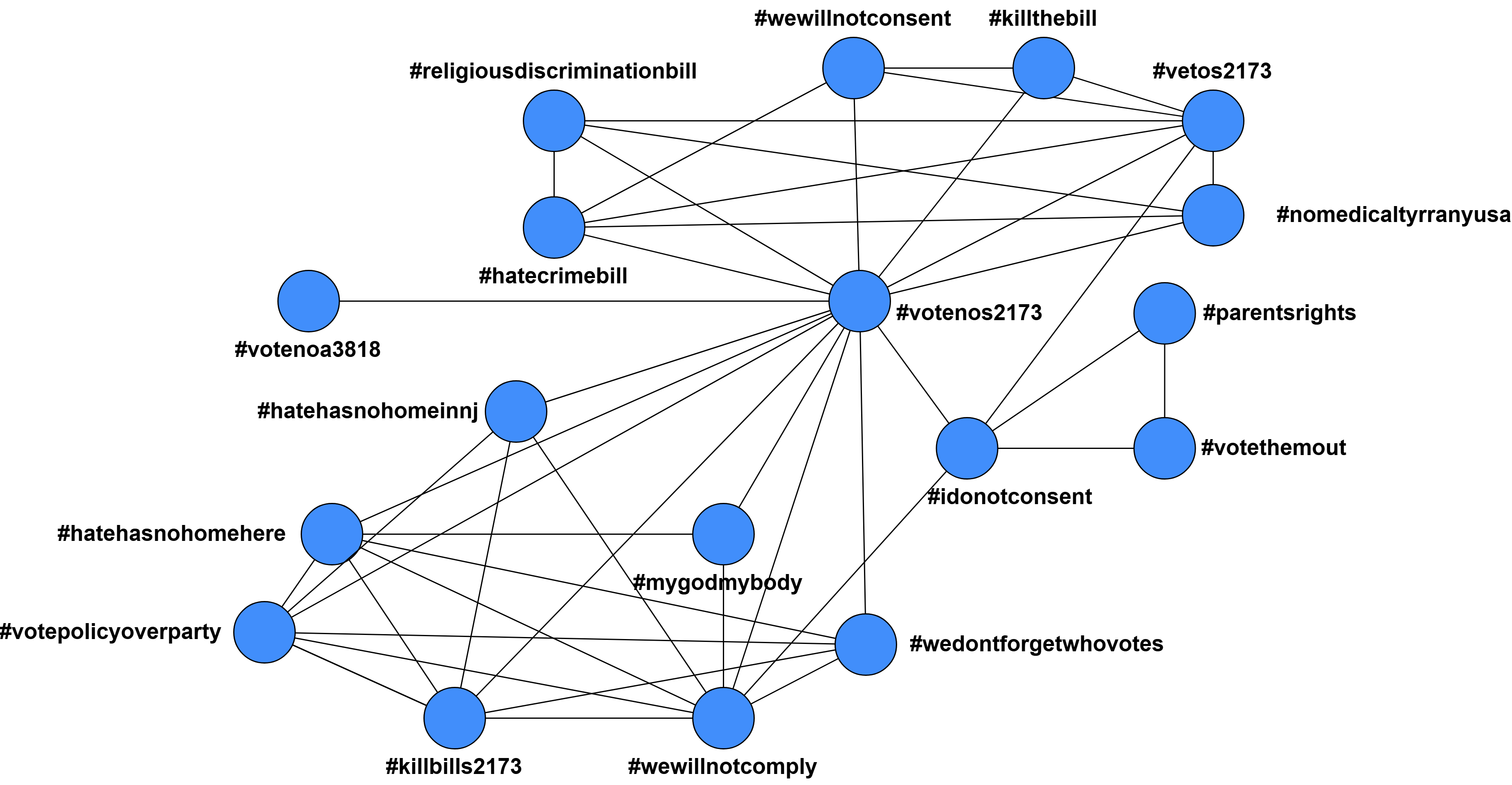}
    \caption{Part\review{ial view} of the blue community from 2019.}
    \label{fig:zoom2-2019}
\end{figure}

The community in blue also stands out. It contains hashtags such as \#mykidsmychoice, \#votethemout, \#firstamendment, \#votenos2173, and \#killbills2173.
The hashtag \#votethemout is probably related to the unacceptance of the government in charge in 2019. The hashtag \#firstamendment prevents the US government from making laws prohibiting religions and their exercise\footnote{Available at: \url{https://www.whitehouse.gov/about-the-white-house/our-government/the-constitution}. Accessed on May 27th, 2023.}. In addition, \#killbills2173 and \#votenos2173 regard Bill S2173\footnote{Available at: \url{https://www.billtrack50.com/BillDetail/966141}. Accessed on May 31st, 2023.}, which intended to eliminate the religious exemption for mandatory childhood vaccination. Therefore, this community combines hashtags related to religion, politics, and vaccination. Fig. \ref{fig:zoom2-2019} shows the major part of the community in blue.

\begin{figure}[!h]
    \centering
    \includegraphics[width=.95\columnwidth]{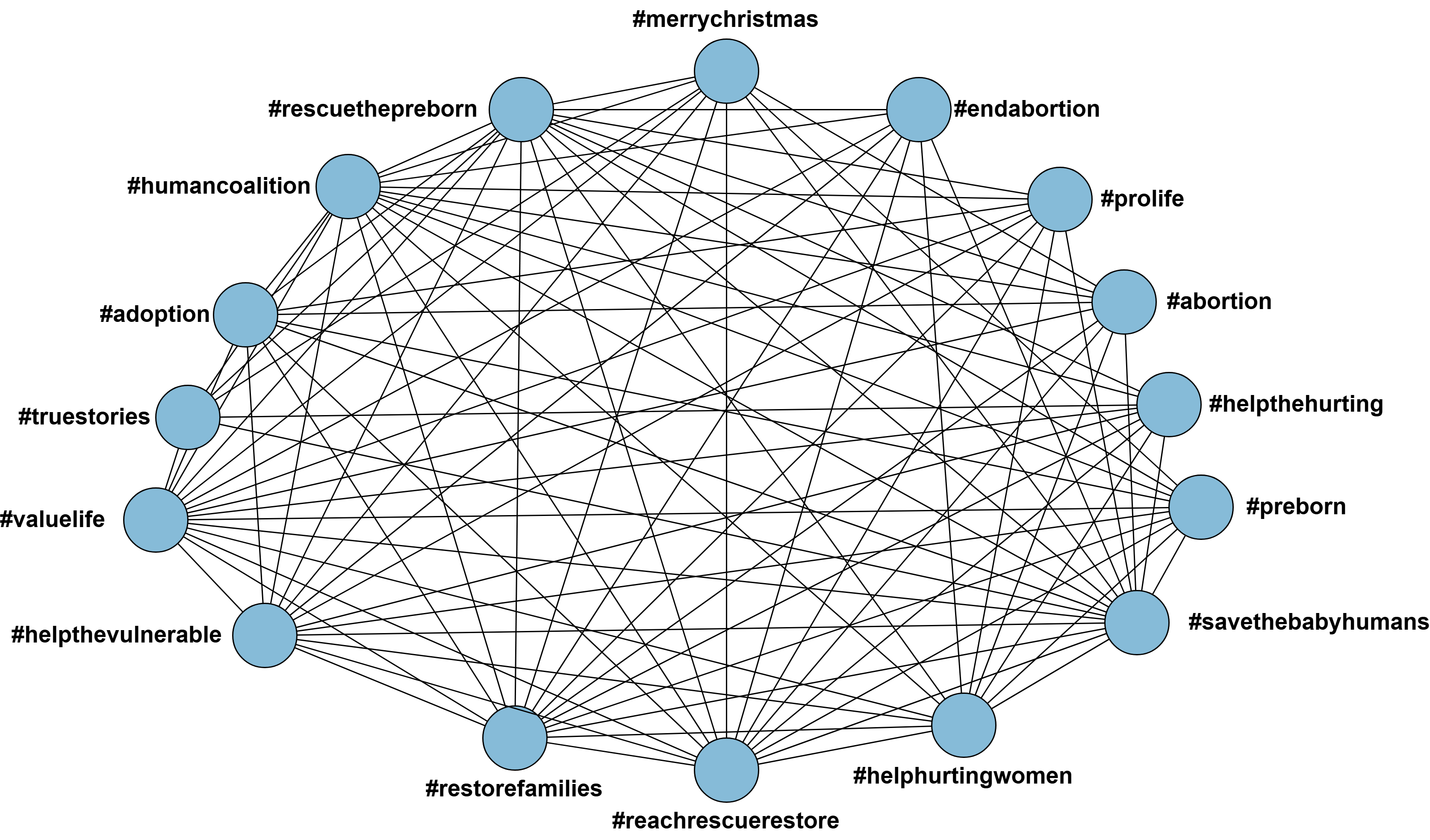}
    \caption{Part\review{ial view} of the blue-gray community from 2019.}
    \label{fig:zoom3-2019}
\end{figure}

\begin{figure}[!h]
    \centering
    \includegraphics[width=.95\columnwidth]{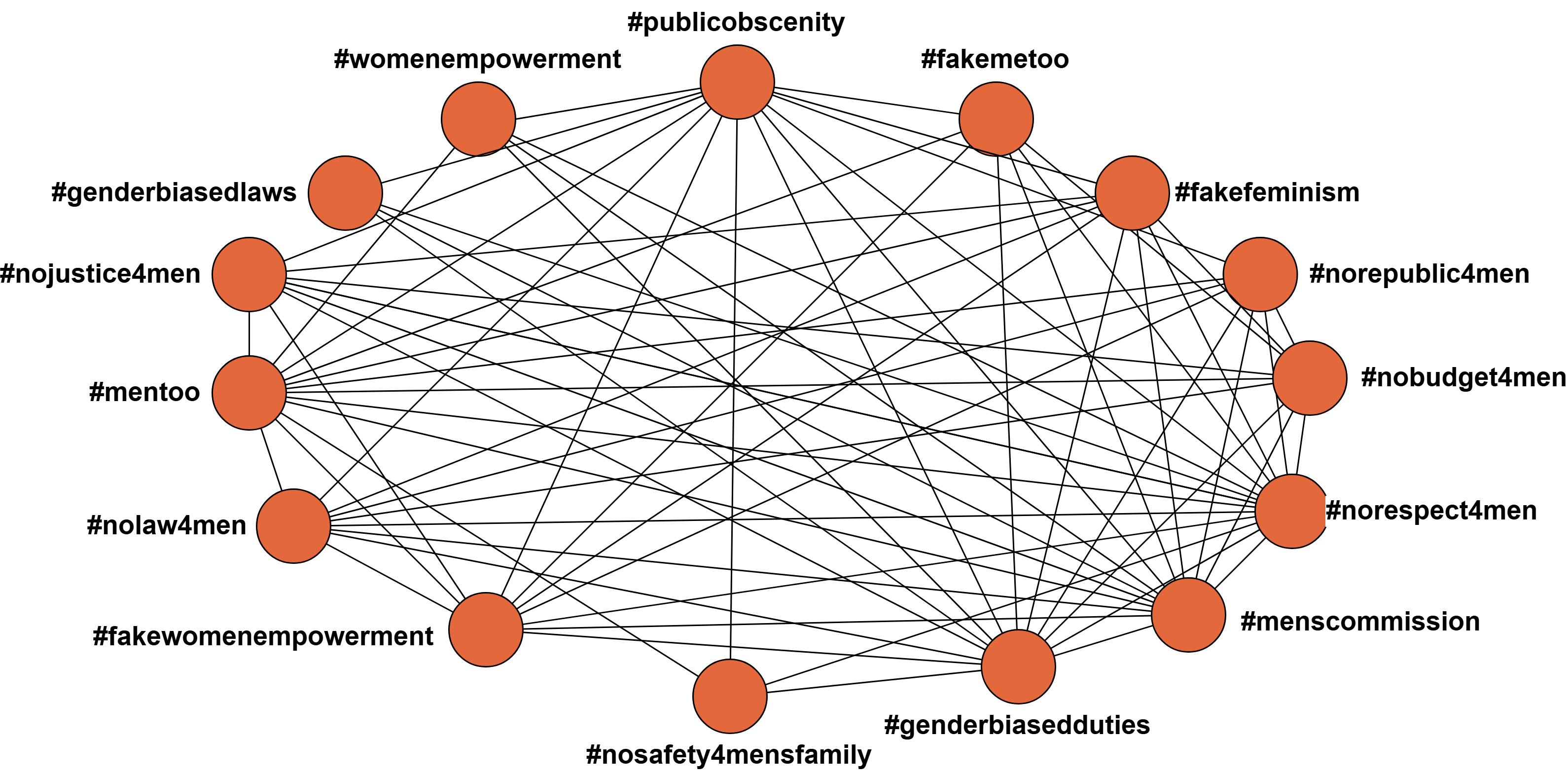}
    \caption{Part\review{ial view} of the dark-orange community from 2019.}
    \label{fig:zoom4-2019}
\end{figure}

The community in blue-gray, i.e., at the center top, considers hashtags such as \#prolife, \#adoption, \#endabortion, \#rescuethepreborn, and \#restorefamilies. This community is undoubtedly anti-abortion. Conversely, the dark-orange-colored community contains hashtags such as \#nojustice4men, \#womenempowerment, \#nolaw4man, \#norepublic4men, and \#genderbiasedlaws. It seems that this community regards feminism or might be considered hate speech. Figs. \ref{fig:zoom3-2019} and \ref{fig:zoom4-2019} shows the blue-gray and the dark-orange communities, respectively.

It is also possible to notice that new communities are growing in 2019, but \review{they} still have a few components. In addition, we see that the \textit{\#mybodymychoice} regards vaccination, feminism, civil rights, and politics in this time slice. The five most frequent hashtags in 2019 and respective frequencies were: \#prochoice (2451), \#womensrights (1384), \#womensrightsarehumanrights (1207), \#abortionisawomansright (916), and \#abortion (905). \review{Table \ref{tab:node_colors2019} summarizes the top 5 communities detected at the end of 2019 with their respective hashtag counts.}

\begin{table}[h!]
\centering
\begin{tabular}{cc}
\hline
\textbf{Context} & \textbf{\# of Hashtags}\\ \hline
{Civil rights, women's rights, and campaign topics} & 86\\
{Anti-abortion} & 28  \\ 
{Religion, politics, and vaccination} & 26  \\
{Feminism/Hate speech} & 16  \\ 
{Vape liberation} & 11  \\ 
\hline 
\end{tabular}
\caption{Top 5 communities and their respective contexts and number of hashtags for 2019.}
\label{tab:node_colors2019}
\end{table}

\subsection{Analysis of 2020}

\begin{figure}[!h]
    \centering
    \includegraphics[width=.9\columnwidth]{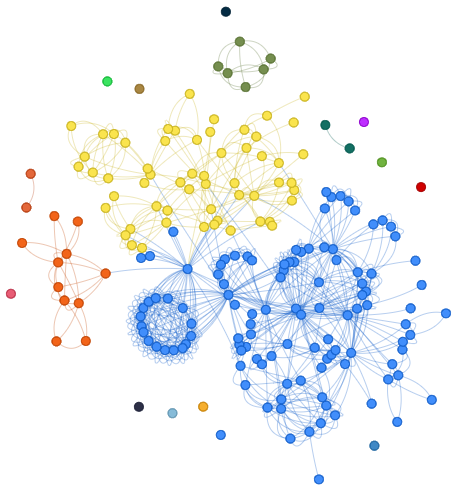}
    \caption{Relationship between hashtags, considering their co-occurrences regarding 2020. The colors represent the communities generated by the Girvan-Newman method. Best viewed in color.}
    \label{fig:2020}
\end{figure}

Fig. \ref{fig:2020} depicts the communities discovered by the Girvan-Newman algorithm considering the hashtags until the end of 2020. Two \review{significant} communities draw attention: those in blue and yellow.

\begin{figure}[!h]
    \centering
    \includegraphics[width=.95\columnwidth]{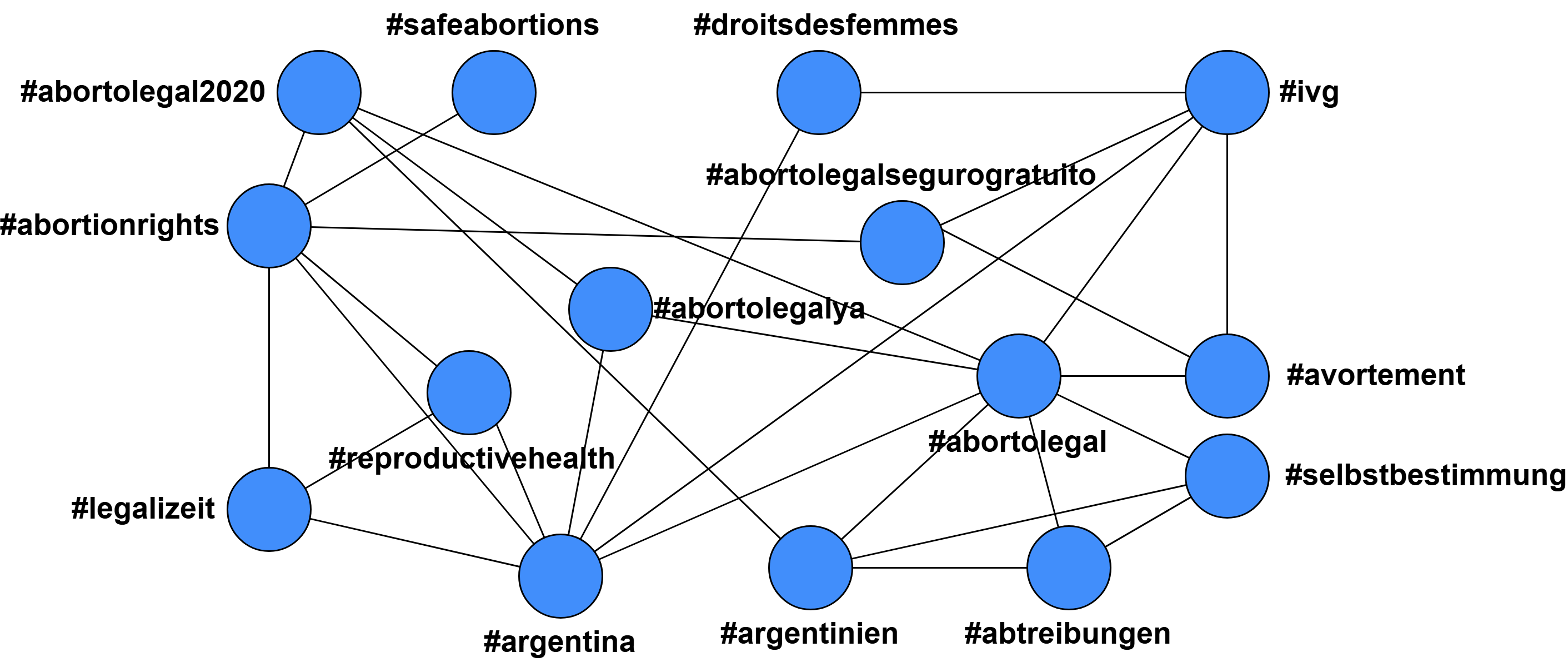}
    \caption{Part of the blue community from 2020.}
    \label{fig:zoom1-2020}
\end{figure}

Exploring the \review{most extensive} community, i.e., the blue-colored community, we see hashtags such as \#abortion, \#abortionrights, \#reproductivehealth, \#womensrights, \#abortionishealthcare, \#abortionwithoutborders, and \#abortionisnotacrime. In addition, hashtags in languages other than English expressing the same stance are present, i.e., \#abortolegal (``legal abortion'' in Portuguese or Spanish), \#avortement (``abortion'' in French), and \#abortolegalya (``legal abortion now'' in Spanish). However, since \#abortion is present in this community, opinions against abortion are present, for example, the hashtags \#rescuethepreborn, \#chooselife, \#savethebabyhumans, and \#endabortions. 
Therefore, this community, depicted in Fig. \ref{fig:zoom1-2020}, combines hashtags in favor and against abortion.

In the yellow community, the terms draw attention. In 2019, the first terms related to regular vaccination appeared. However, in the 2020 analysis, several terms regarding Covid-19 vaccination emerged. 

In November 2019, a new pneumonia-related disease appeared in China, later named Covid-19, after \textit{coronavirus disease} and the year this disease emerged. Due to its ease \review{of} spread and severity, the World Health Organization (WHO) declared Covid-19 a global pandemic in March 2020 \citep{cucinotta2020declares}. Therefore, WHO suggested measures to prevent the Covid-19 spread, such as wearing masks, keeping distance from other people, and washing hands\footnote{Available at: \url{https://www.who.int/westernpacific/emergencies/covid-19/information/transmission-protective-measures}. Accessed on May 30th, 2023.}. 
The US Food and Drugs Association (FDA) authorized the first Covid-19 vaccine for emergency use in December 2020\footnote{Available at: \url{https://www.fda.gov/news-events/press-announcements/fda-takes-key-action-fight-against-covid-19-issuing-emergency-use-authorization-first-covid-19}. Accessed on May 30th, 2023.}.

\begin{figure}[!h]
    \centering
    \includegraphics[width=.95\columnwidth]{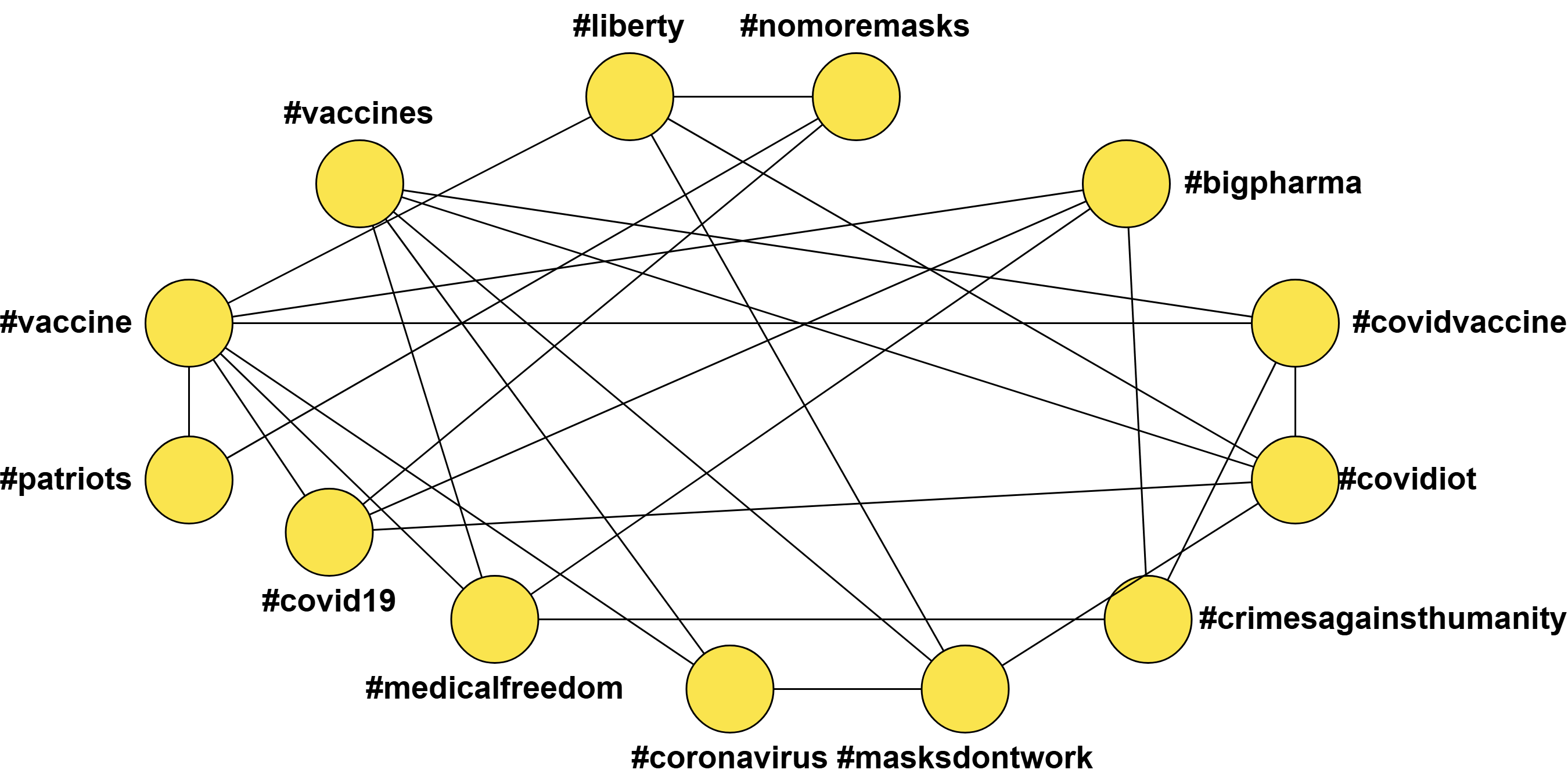}
    \caption{Part of the yellow community from 2020.}
    \label{fig:zoom2-2020}
\end{figure}

Thus, 2020 was a year of changes in daily life, and the hashtags demonstrate that. Hashtags such as \#nomoremasks, \#masksdontwork, \#lockdownsdontwork, \#standup, \#medicalfreedom, \#covidiot, \#pharma, and \#covid19 appeared, demonstrating opinions against the masks, vaccines, and lockdowns. 
Therefore, the yellow community, depicted in Fig. \ref{fig:zoom2-2020}, is majorly related to the Covid-19 pandemic.

\begin{figure}[!h]
    \centering
    \includegraphics[width=.8\columnwidth]{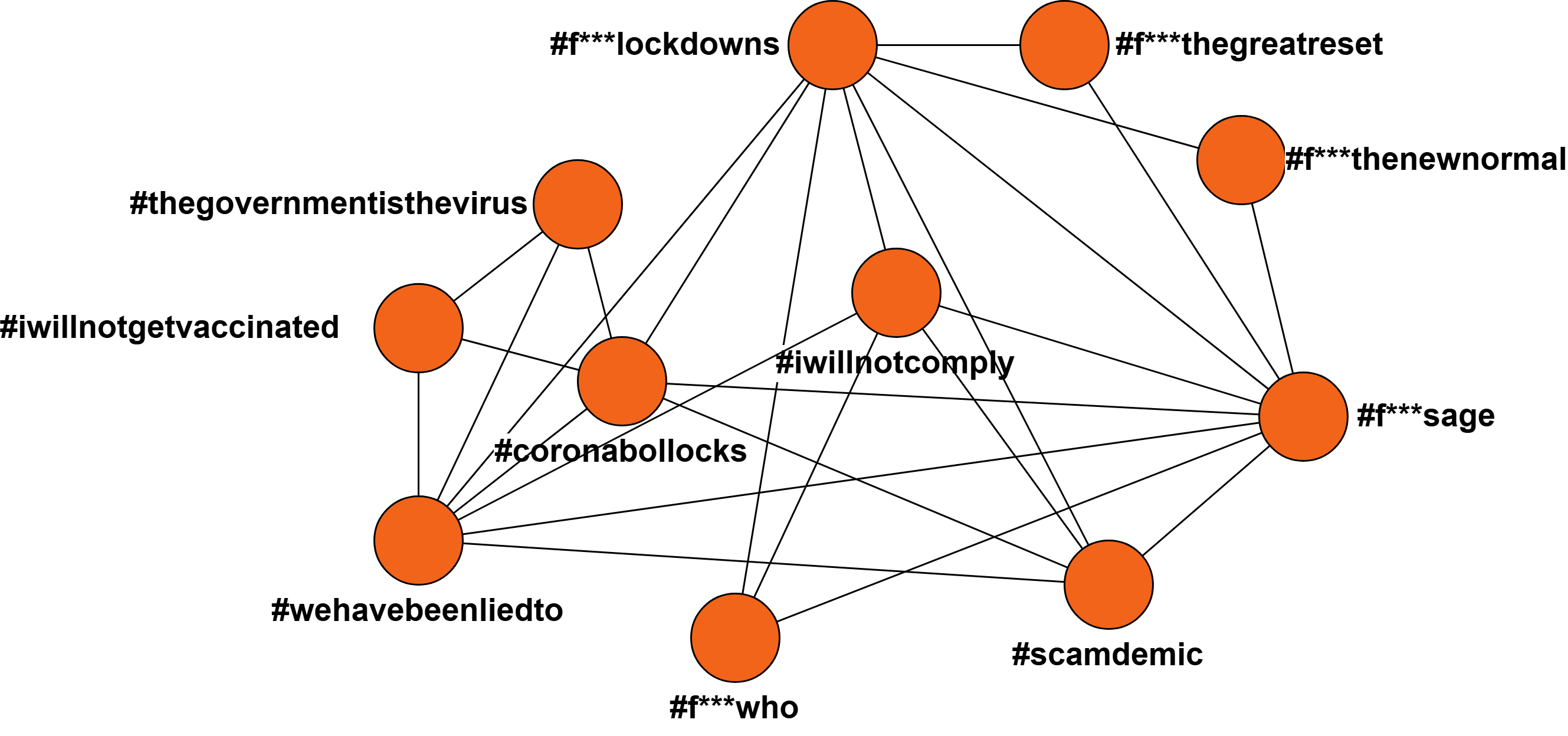}
    \caption{Part of the dark orange community from 2020.}
    \label{fig:zoom3-2020}
\end{figure}

The community in dark orange is also related to the Covid-19 pandemic, most specifically against governmental measures and governments themselves. Some of the hashtags are \#scamdemic (combination of scam and pandemic), \#thegovernmentisthevirus, \#iwillnogetvaccinated, \#iwillnocomply, \#scamdemic, and \#wehavebeenliedto. Fig. \ref{fig:zoom3-2020} depicts this community. We suppressed a few hashtags due to the existence of swearwords.

\begin{figure}[!h]
    \centering
    \includegraphics[width=.7\columnwidth]{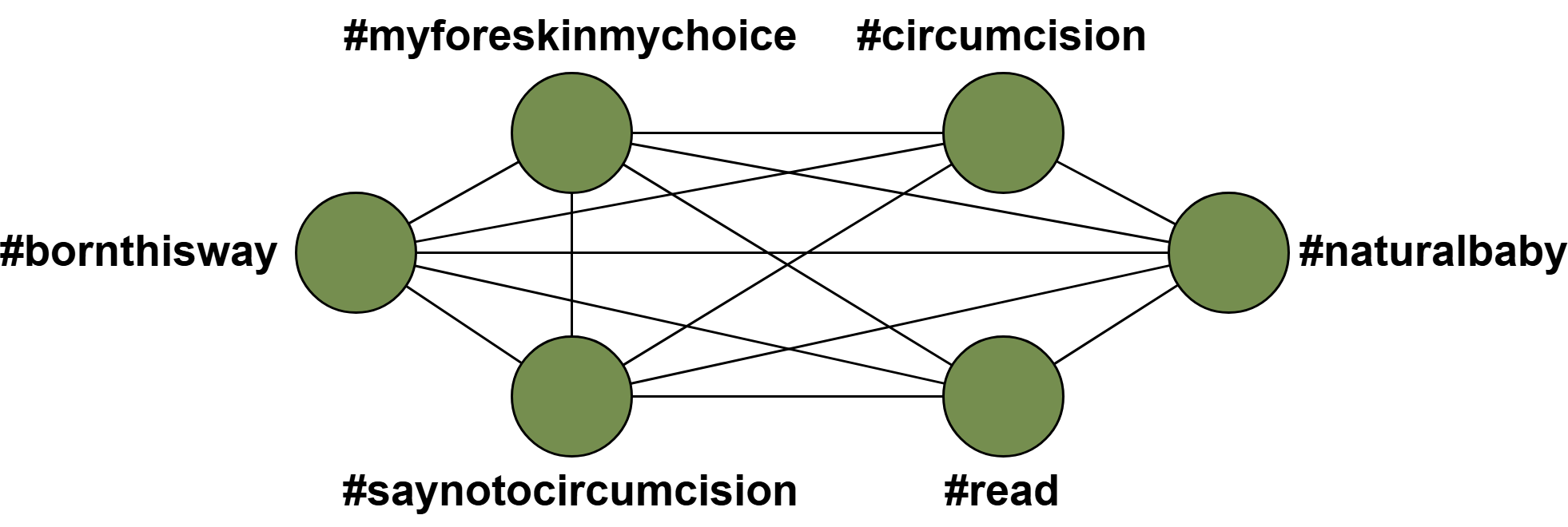}
    \caption{The green community from 2020.}
    \label{fig:zoom4-2020}
\end{figure}

Finally, the community in green regards circumcision, which regards the foreskin removal through surgery. Some of the hashtags are \#myforeskinmychoice, \#circumcision, \#saynotocircumcision, \#bornthisway, and \#naturalbaby. The hashtags demonstrate opinions against circumcision. We can also see new hashtags without communities appearing. Fig. \ref{fig:zoom4-2020} depicts this community.

In summary, although there are hashtags related to the original aim of \#mybodymychoice, it is possible to see that the hashtag \textit{\#mybodymychoice} shifted to topics other than abortion and women's bodily autonomy. Covid-19-related hashtags emerged in 2020. The five most frequent hashtags in 2020 and respective frequencies were: \#prochoice (134), \#abortion (64), \#strajkkobiet (54), \#votenos2173 (53), and \#prolife (50).
\review{Table \ref{tab:node_colors2020} summarizes the top 5 communities detected at the end of 2020 with their respective hashtag counts and contexts.}

\begin{table}[h!]
\centering
\begin{tabular}{cc}
\hline
\textbf{Context} & \textbf{\# of Hashtags}\\ \hline
{Abortion, and women's rights} & 111\\
{Vaccination, and Covid-19 vaccination} & 57  \\ 
{Against governmental measures regarding Covid-19} & 11  \\
{Circumcision} & 6  \\ 
{Not possible to infer} & 2  \\ 
\hline 
\end{tabular}
\caption{Top 5 communities and their respective contexts and number of hashtags for 2020.}
\label{tab:node_colors2020}
\end{table}

\subsection{Analysis of 2021}

\begin{figure}[!h]
    \centering
    \includegraphics[width=\columnwidth]{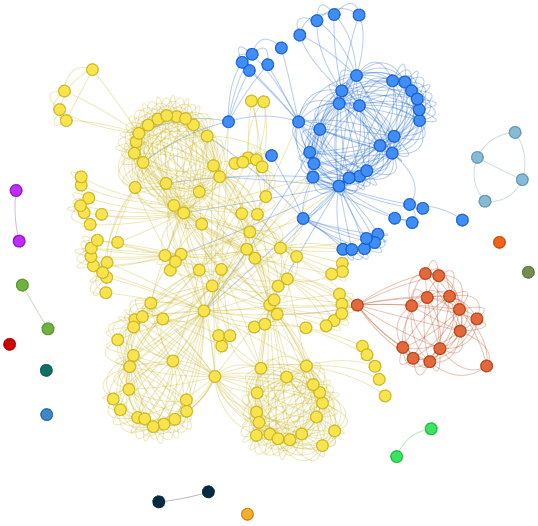}
    \caption{Relationship between hashtags, considering their co-occurrences regarding 2021. The colors represent the communities generated by the Girvan-Newman method. Best viewed in color.}
    \label{fig:2021}
\end{figure}

\begin{figure}[!h]
    \centering
    \includegraphics[width=.65\columnwidth]{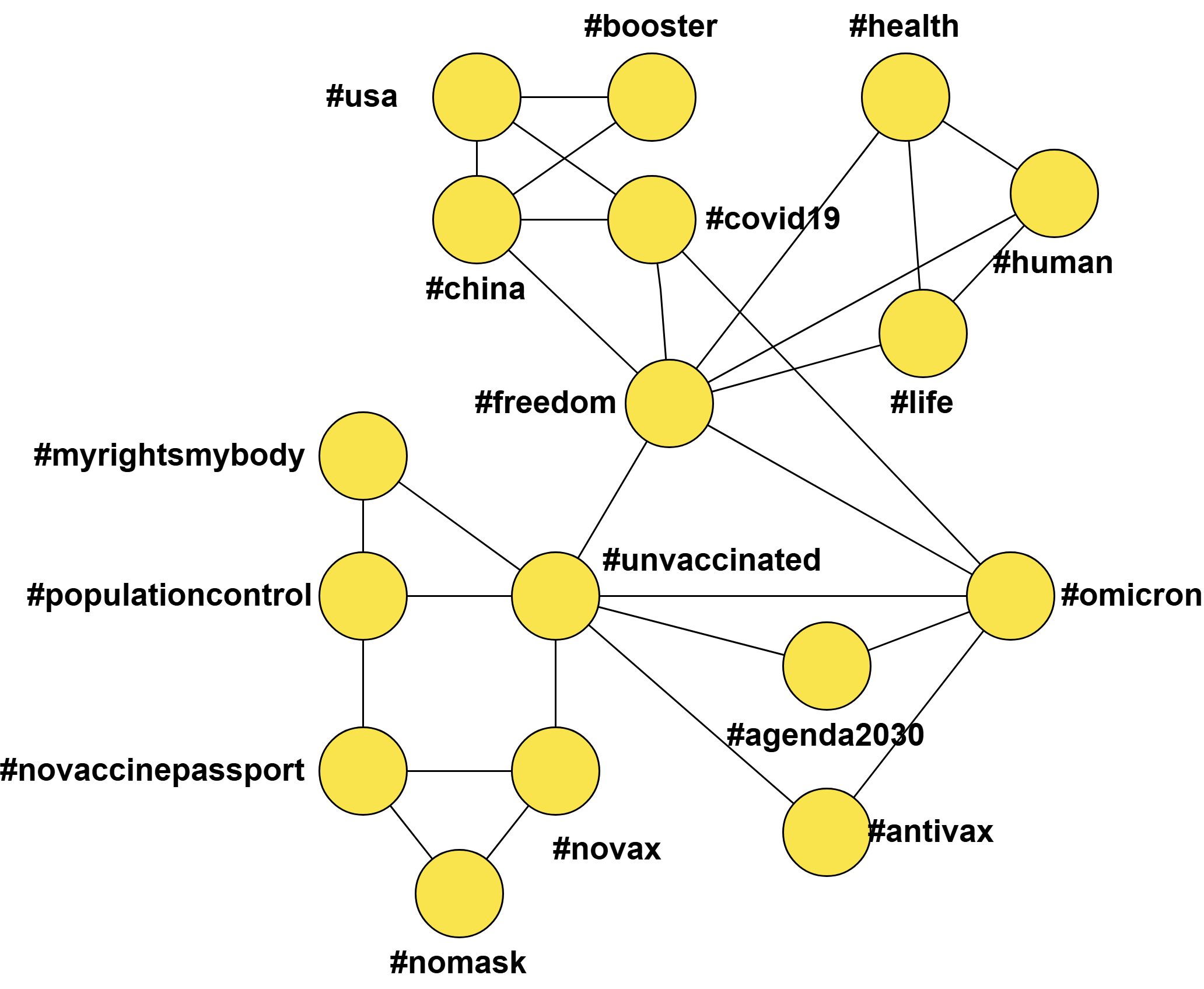}
    \caption{Part\review{ial view} of the yellow community from 2021.}
    \label{fig:zoom1-2021}
\end{figure}

The hashtags relationships collected at the end of 2021 are depicted in Figure \ref{fig:2021}. The community in yellow contains hashtags such as \#dictator, \#eucovidcertificate, \#antivax, \#novaccinepassportanywhere, \#fighttyrany, \#freedomofchoice, \#novax, and \#covidscam. We see that most hashtags are against vaccination and vaccine passport, and in favor of having the choice to have the vaccine or not. Although these hashtags essentially express opinions against vaccination or its governmental control, some hashtags in favor of vaccination also appear. For example, \#getvaccinated and \#getvaccinatednow appear among the contrary hashtags. 
Considering the analysis of the studied period, i.e., between 2018 and 2022, this is the first time the \review{most extensive} community is not related to abortion and women's bodily autonomy. Fig. \ref{fig:zoom1-2021} shows part of the yellow community.

\begin{figure}[!h]
    \centering
    \includegraphics[width=\columnwidth]{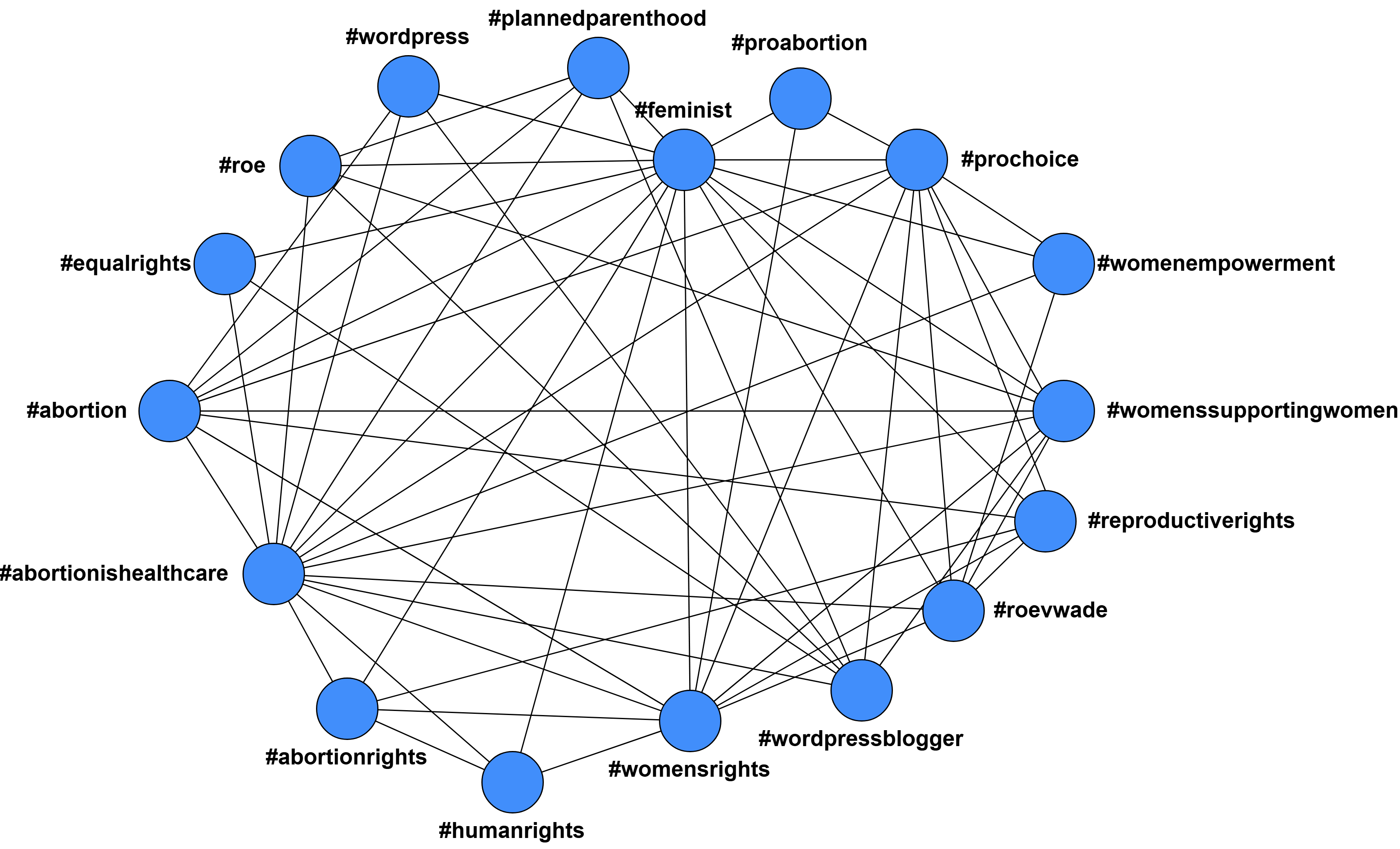}
    \caption{Part of the blue community from 2021.}
    \label{fig:zoom2-2021}
\end{figure}

The blue community considers hashtags from different topics but majorly regards abortion and women's bodily autonomy. Some of the hashtags are: \#abortionishealthcare, \#womensrights, \#plannedparenthood, \#proabortion, and \#reproductiverights. In addition, hashtags such as \#roevwade and \#roe appear. These hashtags regard a decision taken by the US Supreme Court in favor of women choosing to have an abortion, which previously was criminalized\footnote{Available at: \url{https://supreme.justia.com/cases/federal/us/410/113/}. Accessed on May 30th, 2023.}. Unrelated hashtags, such as \#animalrights, \#storytelling, \#writers, and \#wordpress also appear. Fig. \ref{fig:zoom2-2021} partially depicts the blue community.

\begin{figure}[!h]
    \centering
    \includegraphics[width=.6\columnwidth]{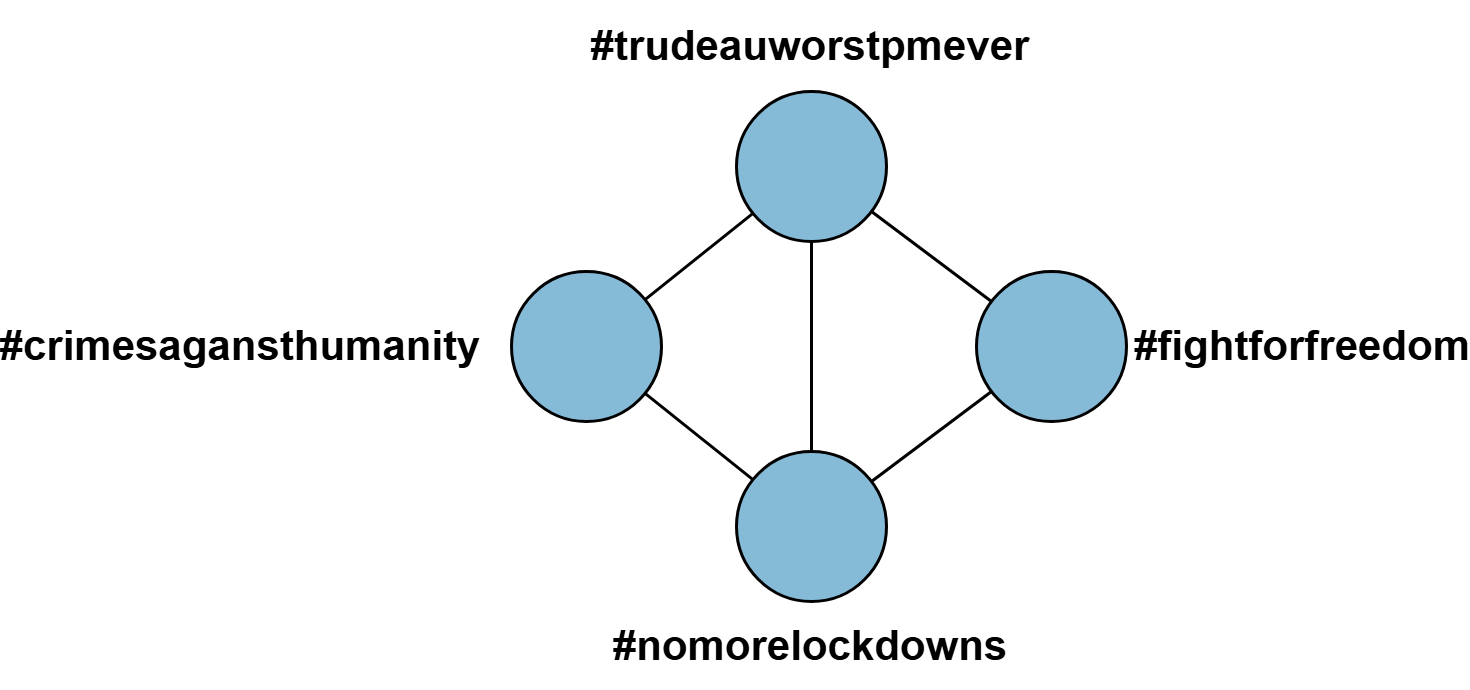}
    \caption{Blue-gray community from 2021.}
    \label{fig:zoom4-2021}
\end{figure}

The community in dark orange demonstrates opinions against the Covid-19 vaccine, specifically Sinovac, with the hashtags \#sinopharm, \#hypocrite, \#corrupt, and \#sinovac. The community in blue-gray considers opinions against lockdowns and the Canadian Prime Minister in charge, Justin Trudeau. The hashtags in this community are \#fightforfreedom, \#trudeauworstpmever, \#nomorelockdowns, and \#crimesagainsthumanity. Both communities are depicted, respectively, in Figs. \ref{fig:zoom4-2021} and \ref{fig:zoom4-2021}.

\begin{figure}[!h]
    \centering
    \includegraphics[width=.9\columnwidth]{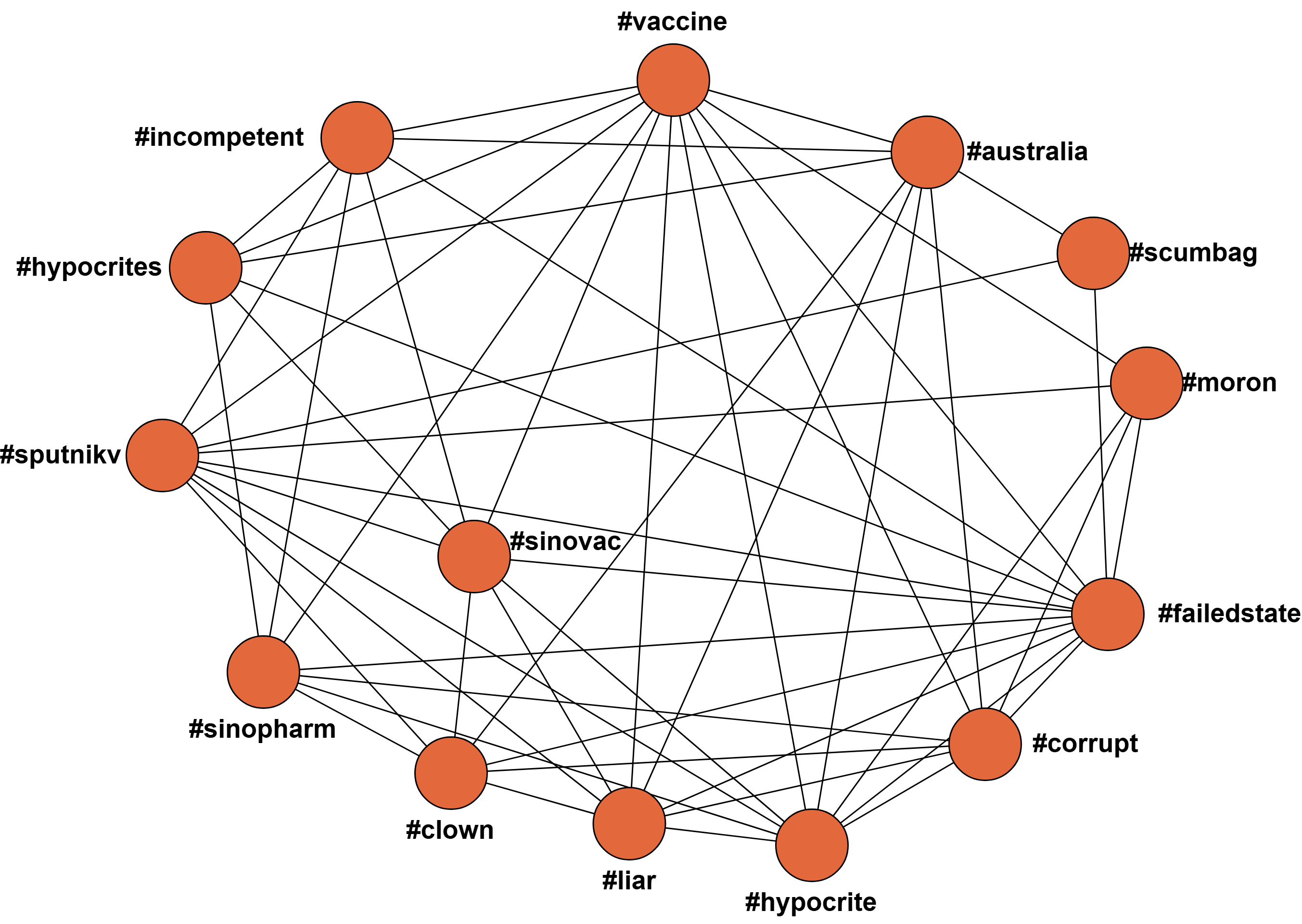}
    \caption{Dark-orange community from 2021.}
    \label{fig:zoom3-2021}
\end{figure}

In summary, the analysis of 2021 highlights the first time in the analyzed period that the biggest detected community does not relate to abortion and women's bodily autonomy. It shows that, at least temporarily, the hashtag \textit{\#mybodymychoice} drifted to vaccination and Covid-19-related topics.
The five most frequent hashtags in 2021 and respective frequencies were: \#novaccinepassports (3694), \#prochoice (2736), \#novaccinepassportsanywhere (2605), \#freedom (2524), and \#novaccinemandates (2305).
\review{Table \ref{tab:node_colors2021} summarizes the top 5 communities detected at the end of 2021 with their respective hashtag counts and contexts.}

\begin{table}[h!]
\centering
\resizebox{\linewidth}{!}{
\begin{tabular}{cc}
\hline
\textbf{Context} & \textbf{\# of Hashtags}\\ \hline
{Covid-19, vaccination, and governmental measures} & 123\\
{Abortion, and women's bodily autonomy} & 45  \\ 
{Against Covid-19 vaccines} & 14  \\
{Against lockdown and Trudeau (Canadian Prime Minister in 2021)} & 4  \\ 
{Not possible to infer} & 2  \\ 
\hline 
\end{tabular}}
\caption{Top 5 communities and their respective contexts and number of hashtags for 2021.}
\label{tab:node_colors2021}
\end{table}

\subsection{Analysis of 2022}

\begin{figure}[!h]
    \centering
    \includegraphics[width=\columnwidth]{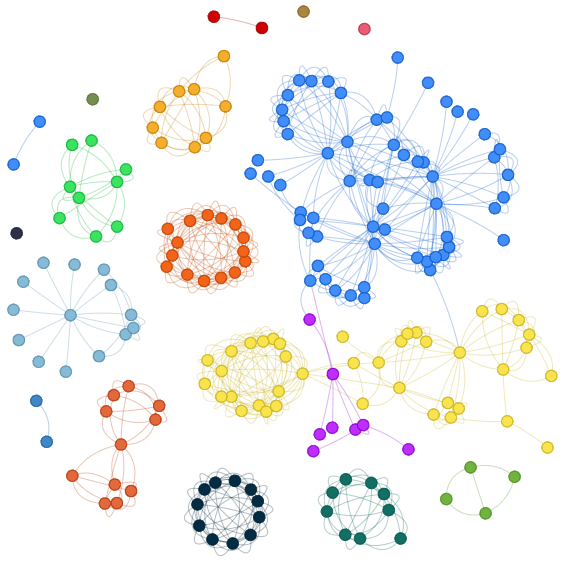}
    \caption{Relationship between hashtags, considering their co-occurrences regarding 2022. The colors represent the communities generated by the Girvan-Newman method. Best viewed in color.}
    \label{fig:2022}
\end{figure}

Figure \ref{fig:2022} depicts the communities detected by the Girvan-Newman method at the end of 2022. 
Compared to the previously analyzed years, more communities were detected.

\begin{figure}[!h]
    \centering
    \includegraphics[width=.75\columnwidth]{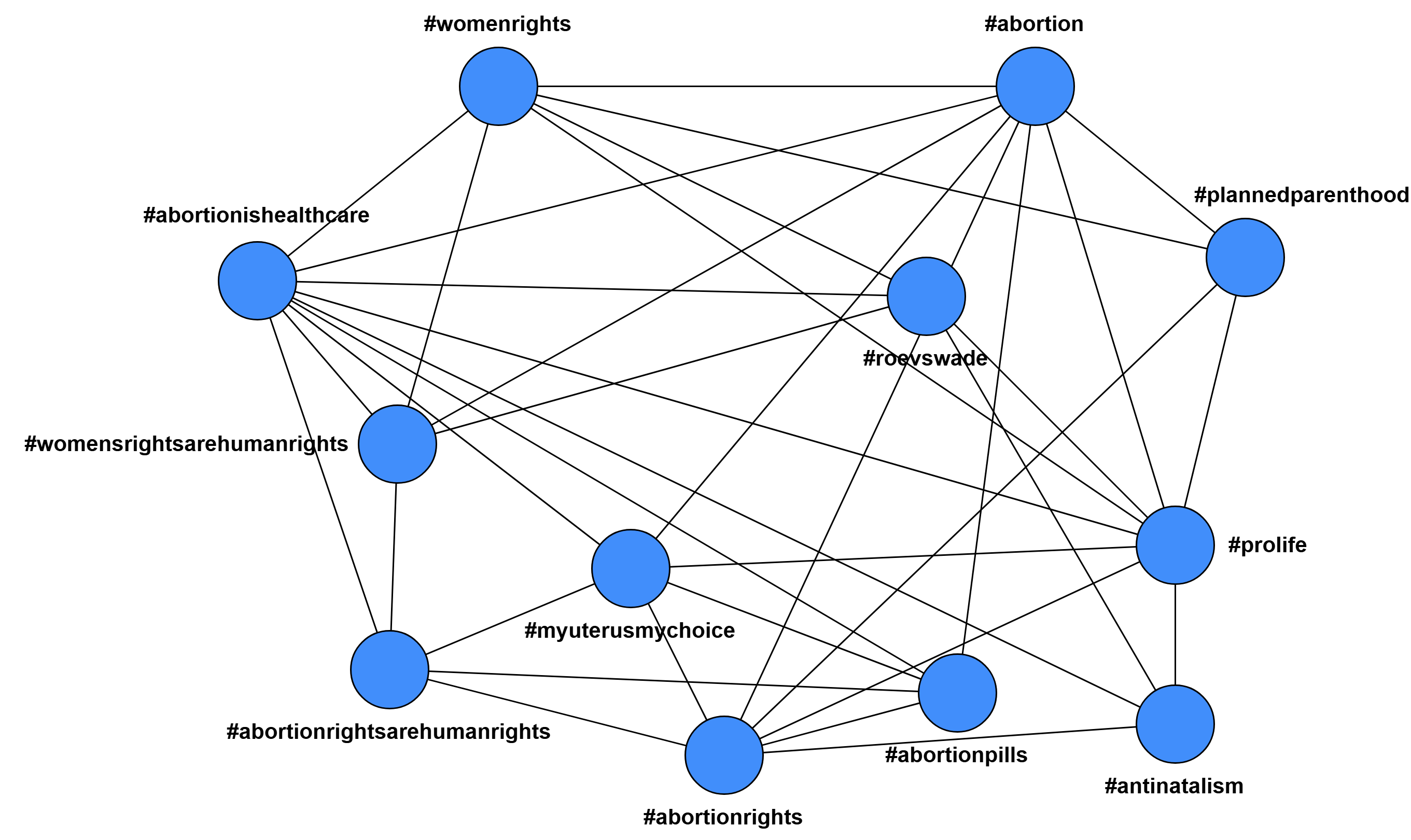}
    \caption{Part\review{ial view} of the blue community from 2022.}
    \label{fig:zoom1-2022}
\end{figure}

Starting with the community in blue, hashtags related to abortion and women's bodily autonomy appear. For example, \#plannedparenthood, \#abortionpills, \#endgunviolence, \#feminism, \#abortionishealthcare, \#abortionrightsarehumanrights, \#myuterusmychoice, \#prochoice, \#prolife, and \#abortion. However, hashtags unrelated to the abortion topic also emerge, such as \#blessed, \#covid, \#endgunviolence, and \#glorytoukraine. The latter two hashtags possibly relate to the Russian-Ukrainian conflict, which started in November 2022. Fig. \ref{fig:zoom1-2022} partially shows the blue community.

\begin{figure}[!h]
    \centering
    \includegraphics[width=.75\columnwidth]{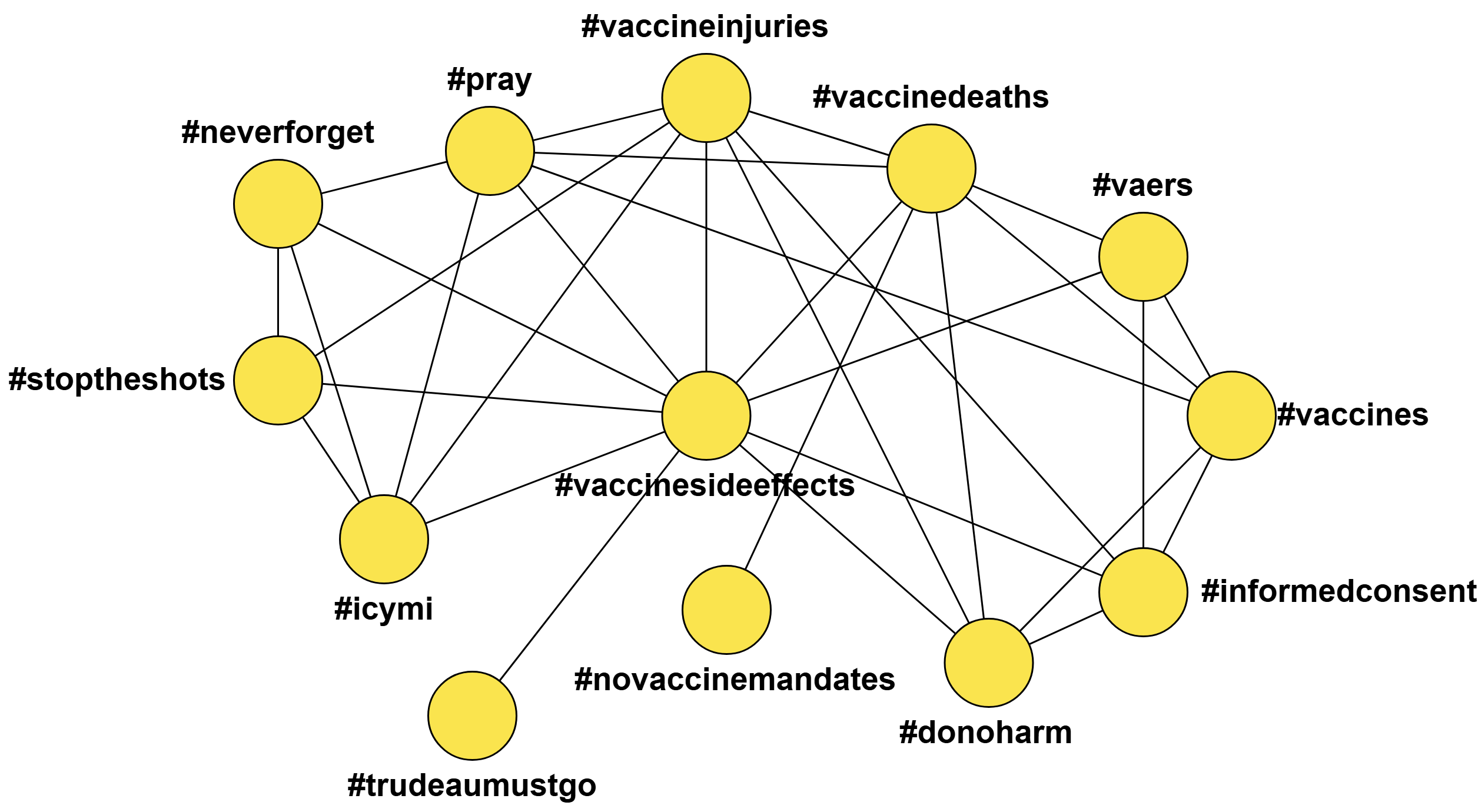}
    \caption{Part\review{ial view} of the yellow community from 2022.}
    \label{fig:zoom2-2022}
\end{figure}

\begin{figure}[!h]
    \centering
    \includegraphics[width=.8\columnwidth]{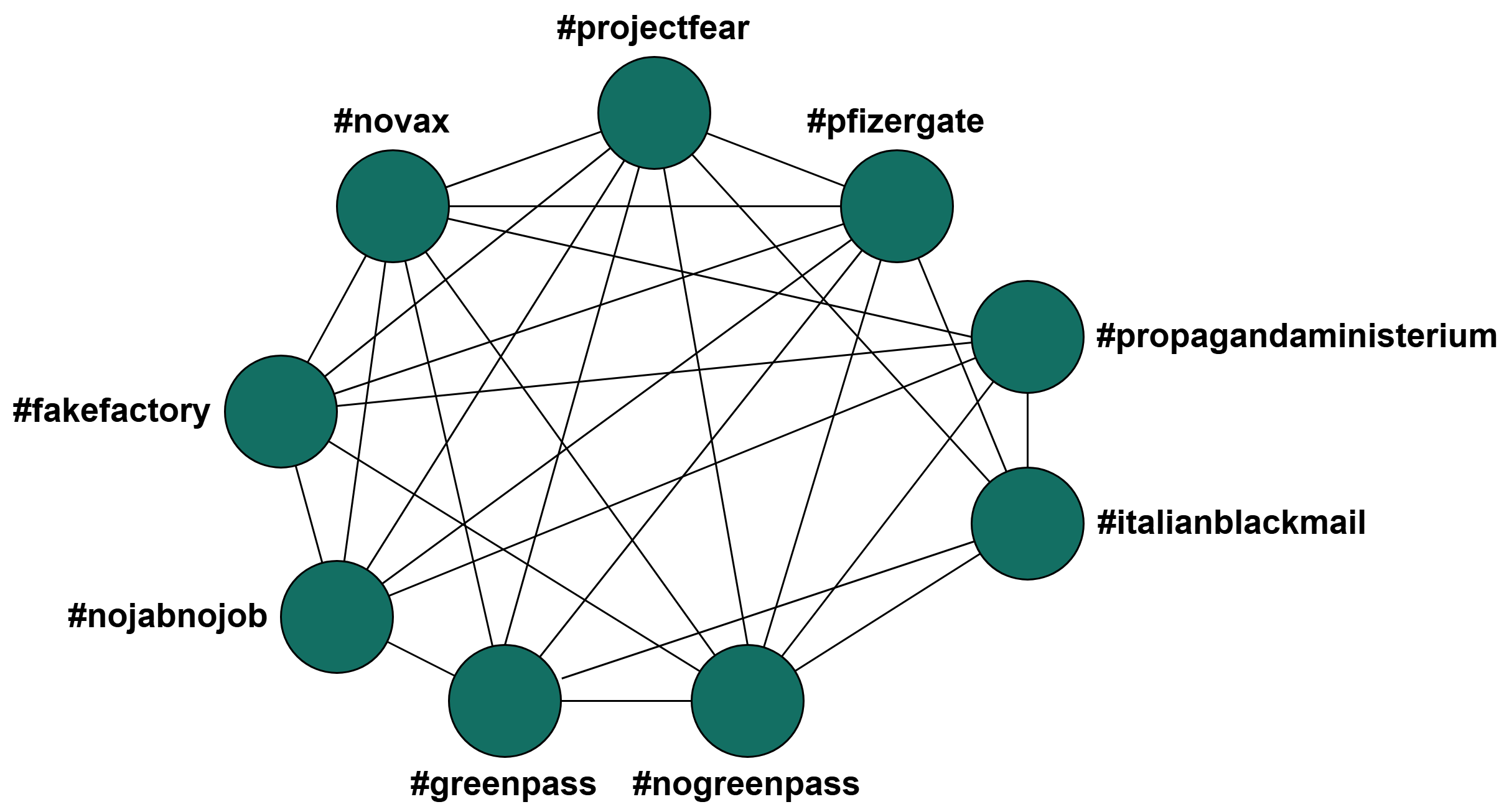}
    \caption{Part\review{ial view} of the dark-green community from 2022.}
    \label{fig:zoom3-2022}
\end{figure}

\begin{figure}[!h]
    \centering
    \includegraphics[width=.9\columnwidth]{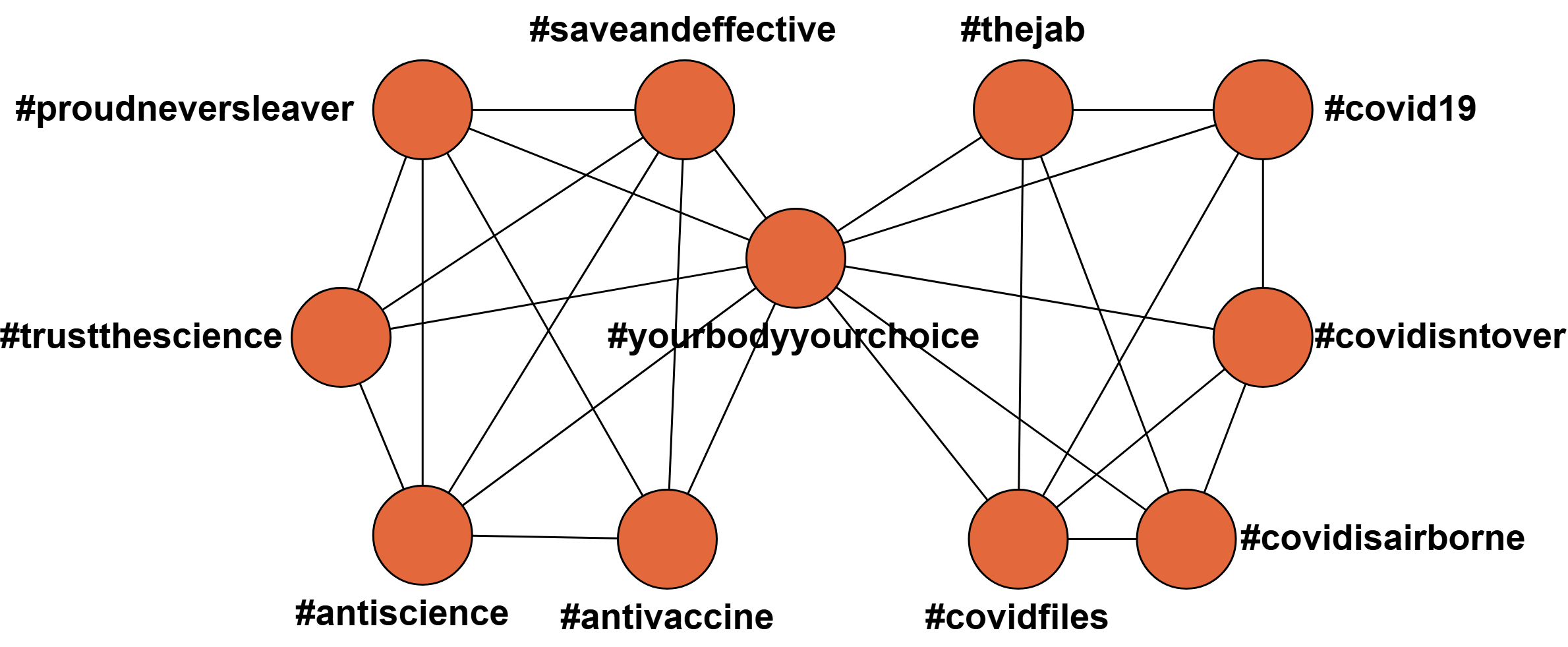}
    \caption{Part\review{ial view} of the dark-orange community from 2022.}
    \label{fig:zoom3.1-2022}
\end{figure}

The community in yellow contains hashtags related to Covid-19. For example, \#novaccinemandates, \#vaccineinjuries, \#vaccinesideeffects, \#informedconsent, \#standupforfreedom, \#vaccinedisaster, \#lockdowns, and \#bigpharma appear in this community. Different from 2021, in 2022, hashtags related to vaccine side effects and injuries appear more frequently than hashtags against vaccination in a Covid-19-related group. The dark-green community also considers Covid-19-related hashtags such as \#greenpass, \#nojabnojob, \#italianblackmail, \#novax, and \#pfizergate. It is closely related to Covid-19 vaccination since Green Pass is the name on which the EU Digital Covid Certificate is known in Italy\footnote{Available at: \url{https://italygreenpass.com/}. Accessed on May 30th, 2023.}. A dark-orange community also emerged containing hashtags related to Covid-19, such as \#yourbodyyourchoice, \#thejab, and \#covidisntover. 
\review{Figs. \ref{fig:zoom2-2022},  \ref{fig:zoom3-2022}, and \ref{fig:zoom3.1-2022} respectively depict partially these communities.}

\begin{figure}[!h]
    \centering
    \includegraphics[width=1\columnwidth]{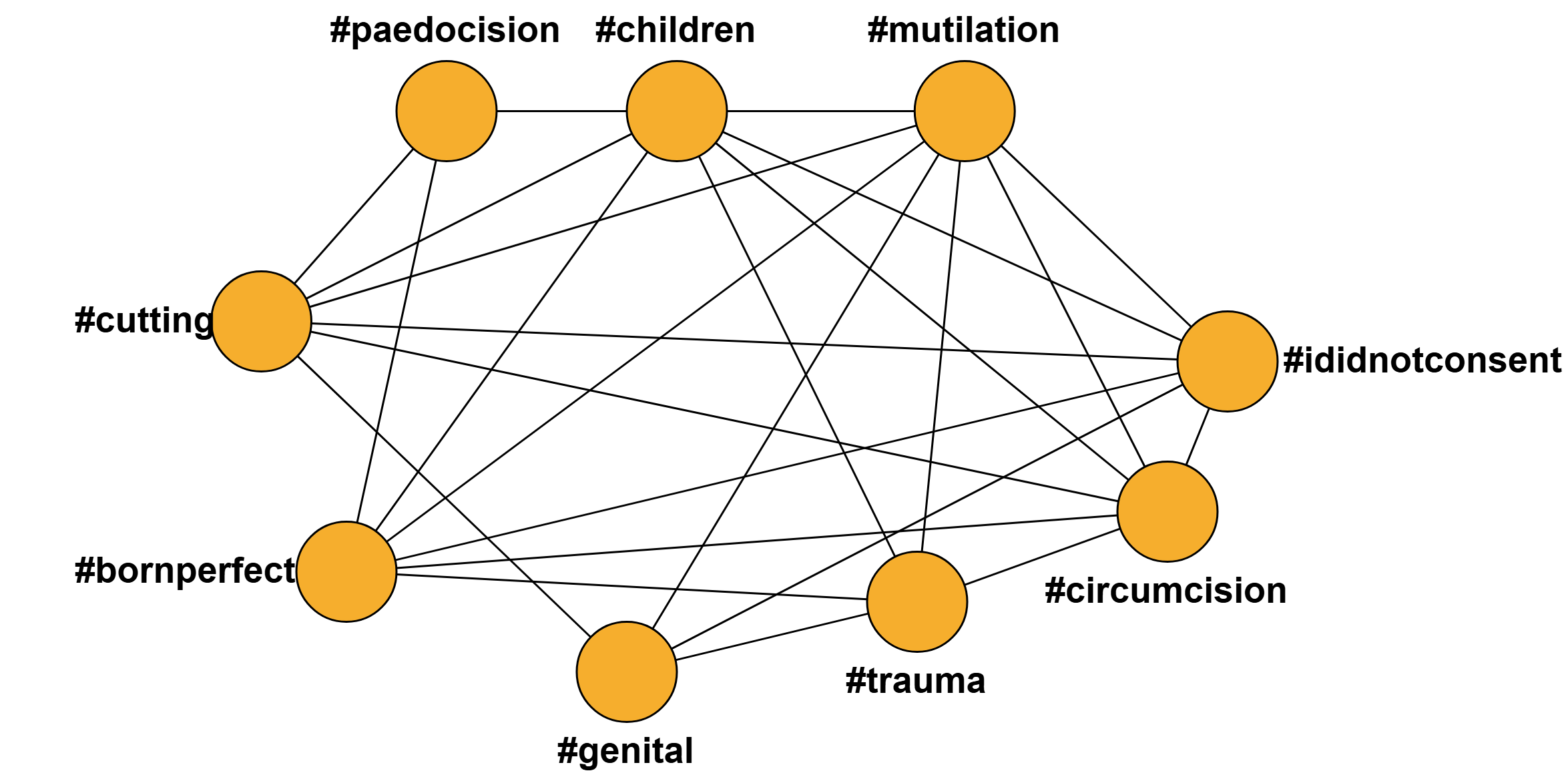}
    \caption{Light orange community from 2022.}
    \label{fig:zoom4-2022}
\end{figure}

The community in blue-gray regards sexual topics, such as \#nudes, which \review{is considered} nudity. Also, the community in dark red is similar to the dark-orange community, having hashtags such as \#braless and \#freethenipples. The community in light orange combines different hashtags, e.g., \#mutilation, \#genital, \#cutting, \#children, and \#circumcision. This last community is related to the circumcision topic, probably against it, since other hashtags, such as \#bornperfect, \#trauma, and \#ididnotconsent are also present, as can be seen in Fig. \ref{fig:zoom4-2022}.

In summary, the \review{most extensive} communities in the analysis of 2022 regard mostly Covid-19, abortion, and women's bodily autonomy-related topics.
Other topics, such as politics and sexual-related, also appear. We also highlight that hashtags related to the Russian-Ukrainian conflict and vaccination side effects emerged. The five most frequent hashtags in 2022 and respective frequencies were: \#roevswade (17719), \#roevwade (11179), \#womensrights (8750), \#prochoice (7987), and \#womensrightsarehumanrights (7961).
\review{Table \ref{tab:node_colors2022} summarizes the top 5 communities analyzed at the end of 2022 with their respective hashtag counts and contexts.}

\begin{table}[h!]
\centering
\begin{tabular}{cc}
\hline
\textbf{Context} & \textbf{\# of Hashtags}\\ \hline
{Women's bodily autonomy} & 62\\
{Covid-19 vaccination} & 40  \\ 
{Sex-related topics} & 15  \\
{Nudity} & 14  \\ 
{Circumcision} & 9  \\ 
\hline 
\end{tabular}
\caption{Top 5 communities and their respective contexts and number of hashtags for 2022.}
\label{tab:node_colors2022}
\end{table}

\section{Conclusion and Future Works}
\label{sec:conclusion}

In this paper, we analyzed the hashtag \textit{\#mybodymychoice} between 2018 and 2022. We generated graphs based on hashtags and their co-occurrences over time and performed annual analyses. The generated graphs went through the Girvan-Newman method for community detection, and the hashtags were discussed to infer topics to the communities and verify whether the hashtag \#mybodymychoice drifted.

We verified that the \#mybodymychoice hashtag was used in topics related to its original use through time: abortion and women's bodily autonomy. 
However, considering the analyzed period, unrelated topics used the aforementioned hashtag. 
For instance, topics such as politics, drugs, Covid-19, and related topics, e.g., vaccination and vaccine passport, used this hashtag, mostly in 2021, where the biggest community found was related to Covid-19. 

More specifically, we aimed to address the research questions presented in Section 
\ref{sec:intro}. 
Considering (RQ1) ``How the hashtag \#mybodymychoice evolved from 2018 to 2022?'', we concluded that the \#mybodymychoice in the period analyzed always showed uses other than its original context, i.e., women's rights, abortion, and bodily autonomy. In 2018, the largest community regarded the hashtag's original context. In 2019, the largest community included hashtags related to vaccination and political campaigns. In 2020, the biggest community regarded the hashtag's original context, using hashtags in languages other than English. Regarding 2021, the yellow community draws attention since it relates to Covid-19, not the original context. In 2022, the biggest community again regards women's rights, abortion, and bodily autonomy. New hashtags related to the Russian-Ukrainian conflict also appear. 

Regarding (RQ2) ``What were the most significant hashtag drifts over time, considering the graph communities detected?'', we concluded that 2021 provided the most significant drift, despite the constant minor drifts of the hashtag \#mybodymychoice. The hashtag drift in 2021 regarded a crucial moment of humanity: the Covid-19 pandemic. Thus, several governmental decisions were questioned by a fraction of society, such as the mandatory vaccination and the use of a document to prove the vaccination, known as the vaccine passport. 

\review{Although our approach provides interesting insights and is suitable for big data scenarios due to the graph creation and management mechanism, it has some limitations. We can mention: (a) the offline community detection step with the Girvan-Newman algorithm, and (b) the need for human analysis for interpreting the communities. Since the limitation (a) corresponds to an offline step, the approach demands 
an interval to be executed. In this paper, we performed community detection yearly.}

In future works, we intend to update and evaluate the described dataset using incremental algorithms for community detection. In addition, \review{we intend to use topic labeling algorithms to infer labels from the hashtags and evaluate the impact of the parameters in the current approach. Furthermore, we also intend to develop an autonomous system for hashtag drift detection}.

\section*{CRediT authorship contribution statement}
\textbf{Cristiano M. Garcia}: Conceptualization, Methodology, Software, Validation, Investigation, Data Curation, Writing - Original Draft, Presentation. \textbf{Alceu de S. Britto Jr.}: Methodology, Validation, Writing - Review \& Editing, Supervision. \textbf{Jean P. Barddal}: Methodology, Validation, Writing - Review \& Editing, Supervision.

\section*{Data Availability}
Under Twitter's policies, the data are publicly available at \url{https://github.com/cristianomg10/temporal-analysis-of-drifting-hashtags-in-textual-data-streams-a-graph-based-application}. \review{In addition, the files with the generated graphs and communities are available in the same repository.}

\section*{Declaration of Competing Interest}
The authors declare that they have no known competing financial interests or personal relationships that could have appeared to influence the work reported in this paper.



\balance
 \bibliography{cas-refs}
\balance




\end{document}